\documentclass[9.5pt,journal,cspaper,compsoc, finalsubmission]{IEEEtran}

\pdfoutput=0

\usepackage[utf8]{inputenc}
\usepackage{graphicx} \usepackage[hyphens]{url}
\usepackage{xcolor}
\usepackage{soul}
\usepackage{listings}
\definecolor{KWColor}{rgb}{0.37,0.08,0.25}
\definecolor{CommentColor}{rgb}{0.133,0.545,0.133}
\definecolor{StringColor}{rgb}{0,0.126,0.941}
\usepackage{tabularx}
\newcolumntype{Y}{>{\centering\arraybackslash}X} 
\usepackage{multirow}
\usepackage[]{hyperref}
\hypersetup{
  colorlinks = true,
    linkcolor = black,
    citecolor = black,
    filecolor = black,
    urlcolor = black,
}
\usepackage{tikz} \usepackage{pgfplots}
\usetikzlibrary{decorations.text}
\usetikzlibrary{calc,trees,positioning,arrows,chains,shapes.geometric,    decorations.pathreplacing,decorations.pathmorphing,shapes,    matrix,shapes.symbols}
\usetikzlibrary{pgfplots.groupplots}
\usepackage{pgfplots}
\usepackage{verbatim}
\makeatletter
\newif\if@restonecol
\makeatother
\usepackage[linesnumbered,ruled,vlined]{algorithm2e}
\usepackage{multirow,bigdelim}
\usepackage{calc}
\usepackage{filecontents}
  
\newenvironment{definition}{}{}
\makeatletter{}

\newcommand{\paperurl}{\url{http://www.abartel.net/permissionmap/}}
\newcommand{\perm}[1]{\mytexttt{#1}}
\newcommand{\method}[1]{{\mytexttt{#1}}}
\newcommand{\class}[1]{\mytexttt{#1}}
\newcommand{\object}[1]{\mytexttt{#1}}
\newcommand{\package}[1]{\mytexttt{#1}}
\def\pdeclared{$P_{d}(app)$}
\def\pinferred{$P_{ifrd}(app)$}
\def\prequired{$P_{req}(app)$}

\newcommand{\chaIntelligent}{CHA-Android}

\newcommand{\sparkEssential}{Spark-Android}
\newcommand{\sparkIntelligent}{Spark-Android}

\newcommand{\androidVersion}{4.0.1}

\newcounter{permsNbr} 

\newcounter{lowPermsNbr} 

\newcounter{nonJavaFrameworkPNbr}
\newcounter{highPermsNbr} 

\newcounter{cppPermsNbr} 

\newcounter{provPermsNbr} 

\newcounter{okPermsNbr} 

\newcommand{\androidFWClasses}{4,071}
\newcommand{\androidFWMethods}{126,660}

\newcommand{\PEPnbr}{6}
 
\makeatletter{}\makeatletter
\DeclareRobustCommand\ttfamily
        {\not@math@alphabet\ttfamily\mathtt
         \fontfamily\ttdefault\selectfont\hyphenchar\font=-1\relax}
\makeatother
\DeclareTextFontCommand{\mytexttt}{\ttfamily\hyphenchar\font=45\relax}
\begin{document}
\title{Static Analysis for Extracting Permission Checks of a Large Scale Framework: The Challenges And Solutions for Analyzing Android}
\author{Alexandre~Bartel,
Jacques~Klein,
Martin~Monperrus, 
and Yves~Le~Traon}
\IEEEcompsoctitleabstractindextext{
\begin{abstract}
\makeatletter{}A common security architecture is based on the protection of certain resources by permission checks (used e.g., in Android and Blackberry).
It has some limitations, for instance, when applications are granted more permissions than they actually need, which facilitates all kinds of malicious usage (e.g., through code injection).
The analysis of permission-based framework requires a precise mapping between API methods of the framework and the permissions they require. 
In this paper, we show that naive static analysis fails miserably when applied with off-the-shelf components on the Android framework. 
We then present an advanced class-hierarchy and field-sensitive set of analyses to extract this mapping. Those static analyses are capable of analyzing the Android framework. They use novel domain specific optimizations dedicated to Android.
 
\end{abstract}
}
\maketitle
\section{Introduction}
	\makeatletter{}
\PARstart{T}{he} security architecture of the mobile operating systems Android and Blackberry as well as other systems such as the Google Chrome browser
extension system, use a similar security model called the permission-based security model \cite{DBLP:conf/ccs/BarreraKOS10}. 
A permission-based security model can be loosely defined as a model in which
1) each application is associated with a set of permissions that allows accessing certain resources\footnote{Contrary to the traditional Unix permission system where permissions are at the level of users, not applications.};
2) permissions are explicitly accepted by users during the installation process and
3) permissions are checked at runtime when resources are requested.
 
In Android, the permission model is embedded into the ``Android framework''. 
The framework exposes an Application Programming Interface (API) that contains classes and methods for developers to interact with the system resources. 
For instance, the API contains a method \method{getGPSLocation}\footnote{simplified view of the API} which gives the current GPS location of the smartphone, if available. 
This API method, and many others, are sensitive with respect to security or privacy.
Consequently, in response to a call to \method{getGPSLocation}, the framework checks that the caller has been explicitly granted the \perm{GPS} permission.
This permission model has an impact on the development process of applications.
To write an application, 
developers must identify, for each API method they use, the permissions that must be declared for the application to work correctly.
They need a mapping between the API methods and the required permissions.
In the case of Android, the mapping is given by the official documentation. 
However, the documentation is not always up-to-date or clear and, consequently, question-and-answers website are full of questions regarding the use of permissions\footnote{e.g.~\url{http://stackoverflow.com/questions/2378607/what-permission-do-i-need-to-access-internet-from-an-android-application/2378619}}.
As a result, developers often either under- or over-estimate the required permissions. 
Missing a permission causes the application to crash. Adding too many of them is not secure. 
In the latter case, injected malware can use those declared, yet unused permissions, to achieve malicious goals.
We call those unused permissions, ``permission gap''.
Any permission gap results in insecure, suspicious or unreliable applications.
\emph{To sum up, having a a clear and precise mapping that links API methods and required permissions is of great value in a permission-based system such as Android.} It enables developers to easily declare the permissions they actually need: not more, not less.
To extract this map, we explore in this paper the use of static analysis to extract the permission checks.
On a framework of the scale and sophistication of Android, naive approaches using off-the-shelf static analysis fail miserably.
This paper discusses the building blocks that must be put together to extract a valuable mapping between API methods and permissions with two kinds of analysis: 
the first kind is based on class hierarchy (CHA) and the second kind leverages a field-sensitive, Andersen \cite{andersen1994program} like module called Spark \cite{lhothend03spark}.
Technically, we describe five components required for extracting permission checks in Android.
The first one is a generic \emph{String analysis}, yet essential for Android where permissions are not static constants but dynamic strings.
The remaining ones are specific to Android. Of those four, the last two components specifically target Spark.
\emph{Service Redirection} redirects call to services to a properly initialized service (Android specific).
\emph{Service Identity Inversion} avoids analyzing irrelevant system calls to services (Android specific).
\emph{Service Initialization} properly initializes services for overcoming null values (Spark specific).
\emph{Entry Points Initialization} initializes all entry point methods and their parameters (Spark specific).
The main difficulty of this research is that, due to the scale and complexity of Android, no building-block yields acceptable result in isolation. 
Eventually, we show that Spark can produce a good mapping of API methods to permissions, and we compare it against the related work \cite{Felt2011a,au2012pscout}.
To sum up, the contributions of this paper are:
\begin{itemize}
\item the empirical demonstration that off-the-shelf static analysis does not address the extraction of permission checks for a framework of the caliber of Android;
\item three static analysis components (generic and Android-specific) to be put together in order to use Class Hierarchy Analysis (CHA) on Android;
\item two static analysis components that allows one to use field-sensitive static analysis (Spark \cite{lhothend03spark}) for analyzing Android's permissions;
\item a comparison of our results against PScout \cite{au2012pscout}, a static analysis designed concurrently with our work and against Felt et al.'s results based on dynamic analysis \cite{Felt2011a};
\item an application of the extracted mapping on two sets of 1421 real Android applications showing that 129 (9\%) applications suffer from a permission gap, i.e., they have more permissions that necessary.
\end{itemize}
This paper is an extension of a short paper published at the International Conference on Automated Software Engineering \cite{bartel:ase2012}.
Those results on this hot topic have been obtained concurrently with other work \cite{Felt2011a,au2012pscout} and explore different paths:
Compared to Felt et al. \cite{Felt2011a}, we use static analysis instead of dynamic analysis. Compared to PScout \cite{au2012pscout}, we go beyond CHA and show that a less naive field-sensitive analysis can also be used.
The reminder of this paper is organized as follows. In Section \ref{sec:manifest-difficult} we explain
why reducing the attack surface is important and present a short study supporting our intuition.
In Section \ref{sec:manifest} we propose a formalization for permission-based software. 
In Section \ref{sec:android} we describe the Android system and its access control mechanisms.
Then, in Section \ref{sec:application-on-android} we extract the permission map from the Android system using static analysis. 
Experiments we conducted and results are presented and discussed in Section \ref{sec:empirical-study}.
In Section \ref{sec:permission-gap} we propose a generic methodology for deriving correct application permission sets.
We present the related work in Section \ref{sec:related-work}.
Finally we conclude the paper and discuss open research challenges in Section \ref{sec:conclusion}. 
 
\section{The Permission Gap Problem}\label{sec:manifest-difficult}
	\makeatletter{}
Let us now detail the permission gap problem introduced in Section 1.
We also present a short empirical study showing that this problem actually happens in practice.
\subsection{Possible Consequence of a Permission Gap}
Let us consider an Android application, $app_{wrong}$, which is able to communicate with external servers since it is granted the \perm{INTERNET} permission.
Moreover, $app_{wrong}$ has declared permission \perm{CAMERA} while it does not use any code related to the camera.
The \perm{CAMERA} permission allows the application to take pictures without user intervention, i.e., the permission gap consists of a single permission: \perm{CAMERA}.
Unfortunately, $app_{wrong}$ uses a native library on which a buffer-overflow exploit has recently been discovered.
As a result, an attacker can execute the code of its choice in the process of $app_{wrong}$ by exploiting the buffer-overflow vulnerability.
The code executed by the attacker in $app_{wrong}$ is granted all permissions defined in $app_{wrong}$, \perm{INTERNET} but also \perm{CAMERA}.
This effectively increases the attacker's privileges.
In this particular example the attacker would be able to (1) write code to use the camera, take a picture and send the picture to a remote host on the Internet 
and (2) execute this code in the target application by exploiting the buffer overflow vulnerability. 
This kind of attack is described in detail by Davi et al. \cite{davi2011privilege}.
On the contrary, if $app_{wrong}$ does not declare \perm{CAMERA}, this attack would not have been possible, and the consequences of the buffer-overflow exploit would have been mitigated.
As noted by Manadhata \cite{Manadhata2011}, reducing the attack surface does not mean no risks, but less risks.
In order to show that this example of misconfigured application is not artificial, we now discuss a short empirical study on the declaration of two permissions on 1,000+ Android applications.
\subsection{Declaration and Usage of Permissions ``camera'' and ``record audio''}
\newcommand{\Mcam}{$M_{\text{CAM}}$}
\newcommand{\Mrec}{$M_{\text{REC\_\-AU\-DIO}}$}
We conducted a short empirical study on 1000+ Android applications downloaded from the Freewarelovers application market\footnote{\url{http://www.freewarelovers.com/android/}}. 
For permissions \perm{CAMERA} and \perm{RE\-CORD\_AU\-DIO}, we grepped the source code of the Android framework to approximate the Then, we computed the list $A$ of all the applications which declare \texttt{CAMERA} or \perm{RECORD\_AUDIO}. 
Next, we took each application $app$ $\in$ $A$ individually and we checked whether the application uses at least one method of If not, it means that $app$ is not using the corresponding permission. 
When this happened, we modified the application manifest that declares the permission and run the application again to make sure that our grepping approximation did not yield false positives.
There are 7/82 applications that declare \perm{CAMERA} while not using it.
Similarly, 3/35 applications declare but do not use \perm{RECORD\_AUDIO} .
Those results confirm our intuition: declared permission lists are not always required, and permission gaps indeed exist. 
Developers would benefit from a tool that automatically infers the set of required permissions and approximates permission gaps.
 
\section{Definitions}\label{sec:manifest}
	\makeatletter{}
Permission-based software is conceptually divided in three layers:
1) the core platform (the operating system) which is able to access all system resources (e.g., change the network policy);
2) a middleware responsible for providing a clean application programming interface (API) to the OS resources and for checking that applications have the right permissions when they want accessing them;
3) applications built on top of the middleware. They have to explicitly declare the permissions they require.
Layers \#2 and \#3 motivate the generic label ``permission-based software''.
Since the middleware also hides the OS complexity and provides an API, it is sometimes called, as in the case of Android, a ``framework''.
Let us now define those terms.
\begin{figure}
\begin{center}
	\makeatletter{}
\pgfdeclarelayer{background}
\pgfdeclarelayer{foreground}
\pgfsetlayers{background,main,foreground}
\tikzstyle{block} = [draw,fill=white!20,minimum width=4em]
\begin{tikzpicture}
  [node distance=.5cm,
  start chain=going below,]
		  \tikzstyle{vertex}=[circle,fill=black!25,minimum size=14pt,inner sep=0pt]
		  \tikzstyle{vInterface}=[block,fill=black!25,minimum size=14pt,inner sep=0pt]
		  \tikzstyle{vPermission}=[diamond,fill=blue!25,minimum size=14pt,inner sep=0pt]
  		\tikzstyle{tuborg}=[decorate]
  		\tikzstyle{tubnode}=[midway, right=2pt]
							  \foreach \name/\x/\y in {fa/1/0, fb/0/1, fc/1/1, fd/2/1, fe/1/2}
		    \node[vertex] (G-\name) at (\x,-\y/1.2) {$\name$};
					  \node[vInterface] (G-e1) at (0,-3/1.2) {$e_{1}$};
		  \node[vInterface] (G-e2) at (1,-3/1.2) {$e_{2}$};
		  \node[vInterface] (G-e3) at (2,-3/1.2) {$e_{3}$};
		  \node[vInterface] (G-in) at (3,-3/1.2) {$e_{4}$};
					  \foreach \from/\to in {fa/fb, fa/fc, fa/e3, fb/e1, fb/fc, fc/fe, fc/fd, fe/fb, fe/fd, fe/e2}
		    \draw[->] (G-\from) -- (G-\to);
\node[right=.4cm of G-fd, text width=2.5cm, font=\scriptsize] (app-permissions) {The application declares permissions $p_1$ and $p_2$};
							  \foreach \name/\x/\y in {f1/0/4, f2/1/4, f3/2/4, f4/0/5, f5/1/5, f6/0/6, f8/3/4, f9/3/5}
		    \node[vertex] (G-\name) at (\x,-\y/1.2) {$\name$};
					  \foreach \name/\x/\y/\n in {ck/1/6/1, ck/4/5/2}
		    \node[draw,vertex,fill=green!15] (G-\name\n) at (\x,-\y/1.2) {$\name_\n$};
					  \foreach \from/\to in {e1/f1, e2/f2, e3/f3, f1/f4, f2/f5, f5/ck1, f4/f6, f4/ck1, f8/f9, f8/ck2, in/f8} 
		    \draw[->] (G-\from) -- (G-\to);
							  \node[vPermission] (G-p3) at (5,-4/1.2) {$p_{3}$};
		  \node[vPermission] (G-p2) at (5,-5/1.2) {$p_{2}$};
		  \node[vPermission] (G-p1) at (5,-6/1.2) {$p_{1}$};
					   \draw[->,dashed] (G-ck1) to (G-p1);
		   \draw[->,dashed] (G-ck2) to (G-p2);
	  \node[vPermission] (Gm-p2) at (1.7,.1) {$p_{1}$};
	  \node[vPermission] (Gm-p2) at (2.2,.1) {$p_{2}$};
		
					  \node[] (G-00) at (0,0) {};
		  \node[] (G-10) at (1.5,0) {};
		  \node[] (G-01) at (0,-3/1.2) {};
		  \node[] (G-11) at (1.5,-3/1.2) {};
			\node[] (G-F00) at (0,-3/1.2) {};
		  \node[] (G-F10) at (4,-3/1.2) {};
		  \node[] (G-F01) at (0,-6/1.2) {};
		  \node[] (G-F11) at (4,-6/1.2) {};
\begin{pgfonlayer}{background}
				        \path (G-00.west |- G-10.north)+(-0.5,0.3) node (a) {};
        \path (G-01.north -| G-11.east)+(+.8,-0.1) node (b) {};
        \path[fill=black!5,rounded corners, draw=black!50, dashed]
            (a) rectangle (b);
				        \path (G-F00.west |- G-F10.south)+(-0.5,0.1) node (a) {};
        \path (G-F01.south -| G-F11.east)+(+0.3,-0.2) node (b) {};
        \path[fill=black!10,rounded corners, draw=black!50, dashed]
            (a) rectangle (b);
\end{pgfonlayer}
						\draw[tuborg, decoration={brace}] let
			    \p1=(G-fa.north), \p2=(G-e1) in
					    ($(-1,\y2)$) -- ($(-1,\y1)$) node[tubnode, left]  {Application$\hspace{.3cm}$};
			\draw[tuborg, decoration={brace}] let
			    \p1=(G-e1), \p2=(G-f6.south) in
					    ($(-1,\y2)$) -- ($(-1,\y1)$) node[tubnode, left]  {Framework$\hspace{.3cm}$};
\end{tikzpicture}
 
\caption{\label{fig:generic-application-framework}A Bird's Eye View of An Application Written on Top of a Permission-based Framework. ($e_n$ are entry points, $f_n$ are functions and methods and $ck_n$ represent checks of permissions $p_n$.)}
\end{center}
\end{figure}
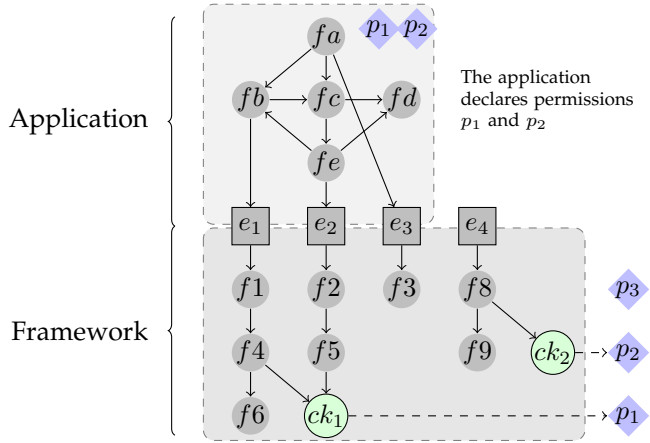
 
	\begin{definition}\textbf{Framework}
	\label{def:framework}  
	A framework  $\mathcal{F}$ is a layer that enables applications to access resources available on the platform. 
	We model it as a bi-partite graph where each node in the set of API method nodes connects a node in the set of resource nodes (this set also contains a 'no resource' node).
	\end{definition}\newline 
\emph{Example}: In Figure \ref{fig:generic-application-framework} the framework is composed of nine methods (four of them being public). Applications access the framework through four API methods. In the case of Android, $\mathcal{F}$ is the Android \androidVersion{} Java Framework
composed of \androidFWClasses{} classes and \androidFWMethods{} methods.
To access a resource, an Android application has to make a method call that goes through $\mathcal{F}$.
	\begin{definition}\textbf{Permission}
	\label{def:pep}
	A permission is a token that an application needs to access a specific resource. 
	\end{definition}\newline
\emph{Example}: In Figure \ref{fig:generic-application-framework}, the application declares two permissions. The framework defines three permissions but only checks two. We make no assumptions on permissions, and we consider them as independent (neither grouped, nor hierarchical).
	\begin{definition}\textbf{Permission-based system}
	\label{def:permission-based-system}
	A permission-based system is composed of  at least one framework, a list of permissions and a list of protected resources. 
Each protected  resource is associated with a fixed list of permissions.
	\end{definition}
	\begin{definition}\textbf{Entry point}
	\label{def:entry-point}
	An entry point of a framework is a method that an application 
	can use (e.g., public or documented). Constructors are also considered as entry points.
	We denote $Entry_{\mathcal{F}}$ as the set of all entry points of $\mathcal{F}$.
	\end{definition}\newline
\emph{Example}: In Figure \ref{fig:generic-application-framework}, there are four entry points (e$_1$ to e$_4$).
An application can call any public method of the framework. 
Some methods accessing system resources (like an account) are protected by one or more permissions.
In the case of Android \androidVersion{}, there are 50,029 entry points.
	\begin{definition}\textbf{Declared permission}
	\label{def:declared-permission}
	A declared permission for an application $app$ is a permission which is in the permission list of $app$.
	The set of all declared permission for an application $app$ is noted \pdeclared.
	\end{definition}\newline
\emph{Example}: In Figure \ref{fig:generic-application-framework}, the application declares $p_{1}$ and $p_{2}$.
In the case of Android, the permissions of an application are declared in a file called \emph{manifest}.
	\begin{definition}\textbf{Required permission}
	\label{def:required-permission}
	A required permission for an application $app$ is a permission associated with a resource that $app$ uses at least once.
	The set of all required permissions for an application $app$ is noted \prequired.
	\end{definition}\newline
\emph{Example}: In Figure \ref{fig:generic-application-framework}, the application requires permission $p_{1}$. 
	\begin{definition}\textbf{Inferred permission}
	\label{def:inferred-permission}
	An inferred permission for an application $app$  is a permission that an analysis technique found to be required for $app$.
	\end{definition}\newline
Depending on the analysis technique used, the inferred permission list may be either an over- or an under- approximation of the required permission list. 
When developers write manifests, they write \pdeclared{} by trying to guess \prequired{} based on documentation and trial-and-errors.
In this paper, we propose to automatically infer a permission list \pinferred{} in order to avoid this manual and error-prone activity.
 
\vspace{-.15cm}
\section{Overview of Android}\label{sec:android}
	\makeatletter{}
\vspace{-.1cm}
This section gives an overview of the architecture of Android in Section \ref{sub:software-stack}.  
We focus on the parts related to permissions in Sections \ref{sub:highLowPermissions} and \ref{subsec:service}.
Other technical details very important for static analysis are discussed in Section \ref{sec:veryTechDetails}.
\vspace{-.25cm}
\subsection{Software Stack}
\label{sub:software-stack}
Android is a system with different layers.  
It consists of a modified Linux kernel, C/C++ libraries, a virtual machine called Dalvik, a Java framework compiled to Dalvik bytecode, and a set of applications.  
Applications for Android are written in Java and compiled into Dalvik bytecode. 
Dalvik bytecode is optimized to run on devices where memory and processing power are scarce.  
An Android application is packaged into an Android package file which contains the Dalvik bytecode, data (pictures, sounds ...) and a metadata file called the ``manifest''.  
\vspace{-.24cm}
\subsection{Android Permissions}
\label{sub:highLowPermissions}
Application vendors define a set of permissions for each application. 
For installing an application, the user has to approve as a whole all the permissions the application's developer has declared in the application manifest.  
If all permissions are approved, the application is installed and receives group memberships.
The group memberships are used to check the permissions at runtime.
For instance, an application \emph{Foo} is given two group memberships \texttt{net\_bt} and \texttt{inet} when
installed with permissions \perm{BLUETOOTH} and \perm{INTERNET},
respectively.  In other terms, the standard Unix ACL is used as an
implementation means for checking permissions.
Android 2.2 defines 134 permissions in the \class{android.Ma\-ni\-fest\-\$per\-mis\-sion} system class, whereas Android 4.0.1 defines 166 permissions.
This gives us an upper-bound on the number of permissions which can be checked in the Android framework.
Android has two kinds of permissions: ``high-level'' and ``low-level'' permissions.
High-level permissions are  only checked at the framework level (that is, in the Java code of the Android SDK).
Android 2.2 declares eight low-level permissions which are either checked in C/C++ native services (\perm{RECORD AUDIO} for instance) or in the kernel 
(e.g., when creating a socket).
In this paper, we focus on the high-level permissions that are only checked in the Android Java framework. 
\subsection{Services and Permissions}
\label{subsec:service}
\begin{figure}
\begin{center}
\resizebox{\columnwidth}{!}{
\makeatletter{}
\begin{tikzpicture} [node distance = 6em]
\newcommand{\myg}{goul}
\newcommand{\mywidth}{.002\columnwidth}
\newcommand{\myl}{li}
\tikzstyle{sensor}=[draw, fill=blue!20, text width=5em, 
    text centered, minimum height=2.5em]
\tikzstyle{ann} = [above, text width=5em]
\tikzstyle{naveqs} = [sensor, text width=6em, fill=red!20, 
    minimum height=12em, rounded corners]
\tikzstyle{zygote} = [draw, text width=6em, 
	text centered, minimum height=15em]
\tikzstyle{app} = [draw, text width=8em, 
	text centered, minimum height=5.6em]
\tikzstyle{service} = [draw, text width=8em, 
 minimum height=6em]
\tikzstyle{bindervoid} = [ minimum width=0.2em, 
	text centered, minimum height=6em]
\tikzstyle{binder} = [draw, minimum width=22em, 
	text centered, minimum height=2em]
\tikzstyle{mystep} = [draw, fill=black!5, circle, inner sep=.1em]
\tikzstyle{systemserver} = [draw, text width=6em, 
	minimum height=15em]
\tikzstyle{servicemanager} = [draw, text width=6em, 
	text centered, minimum height=15em]
\tikzstyle{binderdriver} = [draw, text width=20em, 
	text centered, minimum height=3em, minimum width=24em]
\tikzstyle{stub} = [draw, dashed, rotate=90, minimum height=1.4em, 
	text centered]
\tikzstyle{activity} = [draw, minimum height=13.3em,  minimum width=5em]
\tikzstyle{proxy} = [draw, dashed, rotate=90, minimum height=1.2em, 
	minimum width=1em, text centered]
\def\blockdist{5em}
\def\edgedist{3em}
			\draw node (app1) [app, text width = 10.7em, label=above:\footnotesize Application Code] at (0,0) {};
		\draw node (API1) [draw, text width=10.5em, yshift=-1.3em, label=above:{\small Service Call}] {
r = getSystemService(); \\
p = r.getPassword();
};
			\draw node (bindervoid1) [right of=app1, bindervoid, minimum width=0em] {};
		\draw node (binder1) [color=red, xshift=-.55em, below of=bindervoid1, binder, label=below:{\color{red}\footnotesize Binder}] {};
			\draw node (service1) [right of=bindervoid1, service, text width=8.9em, label=above:{\footnotesize Account System Service}] {getPassword() \{ \\
			{\ }checkPermission();\\
			{\ }return password;\\
			\}};
  \draw node (step1) [mystep] at (API1.west)[xshift=-6, yshift=8] {1};
 \draw node (step2) [mystep] at (API1.west)[xshift=-6, yshift=-8] {2};
 \draw node (step3) [mystep] at (binder1.north)[xshift=8, yshift=-8] {3};
 \draw node (step4) [mystep] at (service1.east)[xshift=4, yshift=7] {4};
 \draw node (step5) [mystep] at (binder1.south)[xshift=30, yshift=7] {5};
  \draw node[circle, minimum width=.25em, draw, fill, color=black, inner sep =0em] (path1start) at (API1.east)[xshift=-15, yshift=-6] {};
 \draw node (path1end) at (service1.west)[yshift=15, xshift=2] {};
 \path [draw, -latex] (path1start) |- (step3) |- (path1end);
  
 \draw node[circle, minimum width=.25em, draw, fill, color=black, inner sep =0em] (path2start) at (service1.south)[xshift=40, yshift=25] {};
 \draw node (path2end) at (API1.south)[xshift=-50, yshift=9] {};
 \path [draw, -latex] (path2start) |- (step5) -| (path2end);
\draw node[] at (path1start.north)[yshift=.005cm] {
\resizebox{\mywidth{}}{!}{ \begin{tikzpicture}
  \tikzstyle{gC} = [circle, inner sep=0cm, outer sep =0cm, fill=green!20];
  \tikzstyle{wC} = [circle, inner sep=0cm, outer sep =0cm, fill=white];
  \tikzstyle{gR} = [rectangle, inner sep=0cm, outer sep =0cm,  fill=green!20];
  \tikzstyle{wR} = [rectangle, inner sep=0cm, outer sep =0cm,  fill=white!20];
  \path (0,0) node[gC, minimum width=1cm] {}
  ++(0, -.5cm) node [wR, minimum width=1cm, minimum height=1cm] {}
  ++(0, .07cm) node [gR, minimum width=1cm, minimum height=.7cm] {}
  +(0, -.15cm) node [gR, minimum width=1cm, minimum height=.7cm, rounded corners] {}
  +(2cm, .7cm) node[] {\myg{}\myl{} \myg{}\myl{}!}
  +(-.7cm, -.05cm) node [gR, minimum width=.25cm, minimum height=.5cm] {}
  +(-.7cm, -.30cm) node [gC, minimum width=.25cm] {}
  +(-.7cm, .20cm) node [gC, minimum width=.25cm] {}
  +(.7cm, -.05cm) node [gR, minimum width=.25cm, minimum height=.5cm] {}
  +(.7cm, -.30cm) node [gC, minimum width=.25cm] {}
  +(.7cm, .20cm) node [gC, minimum width=.25cm] {}
  +(.25cm, -.35cm-.05cm) node [gR, minimum width=.25cm, minimum height=.7cm] {}
  +(.25cm, -.35cm-.40cm) node [gC, minimum width=.25cm] {}
  +(.25cm, -.35cm+.30cm) node [gC, minimum width=.25cm] {}
  +(-.25cm, -.35cm-.05cm) node [gR, minimum width=.25cm, minimum height=.7cm] {}
  +(-.25cm, -.35cm-.40cm) node [gC, minimum width=.25cm] {}
  +(-.25cm, -.35cm+.30cm) node [gC, minimum width=.25cm] {}
  +(.25cm, .7cm) node[wC, minimum width=.1cm] {}
  +(-.25cm, .7cm) node[wC, minimum width=.1cm] {}
  +(.25cm, .9cm) node[gR, rotate=45, minimum width=.1cm] (a) {}
  +(-.25cm, .9cm) node[gR, rotate=-45, minimum width=.1cm] {} ;
\end{tikzpicture}}};
\end{tikzpicture}}
\caption{A Simplified Illustration of the Communication between an Android Application and a Permission Protected Service through the so-called ``Binder''.}
\label{fig:android-system-service}
\end{center}
\end{figure}
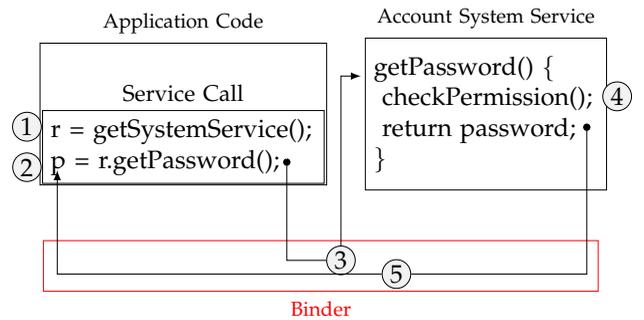

An Android application is made of \emph{components} which can be:
an \emph{Activity} that is a user interface;
a \emph{Service} that runs in background;
a \emph{BroadcastReceiver} (or \emph{Receiver}) that listens for ``intents'' (a kind of
message for inter process communication);
a \emph{ContentProvider} which is a kind of database used to store and share data.
Most permissions are checked at the service level.
Android applications communicate with the operating system using  a special kind of service called \emph{system service}. 
System services are specific services running in a specific scope (called the ``system server'') and allow applications to access system resources (ex: GPS coordinates).  
Those resources may be protected by Android permissions to prevent access by unauthorized applications.
Permission checks associated to services are mostly implemented in Java.
Hence, the scope of our paper consists of analyzing \emph{Android permissions that are enforced in services in the Java framework}.
The impact of this focus is discussed in Section \ref{sec:empirical-study}.
It is important to understand the inner working of system services to devise good static analyses (that will be presented later in Section \ref{sub:intelligentCHA}).
We now describe how the applications communicate with system services. Applications synchronously communicate with system services through a mechanism called  \emph{Binder} as presented in Figure \ref{fig:android-system-service}.  
The first step to communicate with a remote service is to dynamically get a reference (interface) to the service by calling \method{Context.getSystemService()} (step 1 in Figure \ref{fig:android-system-service}). 
The next step is to call a method (method \method{getPassword} from the AccountManager Service in Figure \ref{fig:android-system-service}) from the interface on the object reference \emph{r} (step 2 in Figure \ref{fig:android-system-service}).  
A special component, called ``binder'' is responsible for intercepting  and redirecting that service calls to the remote service that performs the actual computation (steps 3 in Figure \ref{fig:android-system-service}). 
The system service is responsible for enforcing permission checks (step 4 in Figure \ref{fig:android-system-service}).
To check that the caller's application declares the permission in its manifest (Section \ref{sub:software-stack}), 
the service calls one of the methods listed in Table \ref{table:permissionCheckMethods} with the permission to be checked as parameter (not shown in the Figure).
This specific point in the program is called Permission Enforcement Point or PEP.
In Figure \ref{fig:android-system-service}, if the application has the correct permission, the password is returned to the calling application (step 5).
\subsection{Technical Details on Android}
\label{sec:veryTechDetails}
We describe technical details of the Android system.
We leverage this knowledge during static analysis in Section \ref{sec:application-on-android}.
\subsubsection{\label{sub:androidBoot}Android Boot Process}
We describe how Android boots up and what kinds of processes are launched.
It is important to know how to initialize system services when performing precise static analysis with Spark (Section \ref{sub:intelligentSpark}).
If services are not properly initialized, the analysis may be incomplete.
The first program to run on the device is the bootloader which provides support for loading, recovering or updating system images.
The early startup code for loading the Linux kernel is very hardware dependent: it first initializes the environment and only then starts the architecture-independent Linux Kernel C code by jumping to the \method{start\_kernel()} function.
Then, high-level kernel subsystems are initialized (scheduler, system calls, process and thread operations ...) the root filesystem is mounted and the init process is started.
The init process creates mountpoints and mount filesystems, sets up filesystem permissions and starts daemons such as the network daemon, the zygote or the service manager. The zygote is a core process from which new Android processes are forked.
The initialization of zygote starts the system server which in turn initializes system services and managers.
System services include the input manager service and the wifi service.
Managers include the activity manager which handles user interfaces (activities).
Android's boot process indicates that system services and managers are instantiated and initialized at boot time. 
\subsubsection{Android Communication}
\begin{figure}
\begin{center}
\resizebox{\columnwidth}{!}{
\makeatletter{}
\begin{tikzpicture} [node distance = 0em]
\tikzstyle{sensor}=[draw, fill=blue!20, text width=5em, 
    text centered, minimum height=2.5em]
\tikzstyle{ann} = [above, text width=5em]
\tikzstyle{naveqs} = [sensor, text width=6em, fill=red!20, 
    minimum height=12em, rounded corners]
\tikzstyle{zygote} = [draw, fill=white!20, text width=6em, 
	text centered, minimum height=15em]
\tikzstyle{app} = [draw, fill=white!20, text width=8em, 
	text centered, minimum height=5.6em]
\tikzstyle{service} = [draw, fill=white!20, text width=8em, 
 minimum height=6em]
\tikzstyle{bindervoid} = [ minimum width=0.2em, 
	text centered, minimum height=6em]
\tikzstyle{binder} = [draw, fill=white!20, minimum width=22em, 
	text centered, minimum height=2em]
\tikzstyle{mystep} = [draw, fill=black!5, circle, inner sep=.1em]
\tikzstyle{systemserver} = [draw, fill=white!20, text width=6em, 
	minimum height=15em]
\tikzstyle{servicemanager} = [draw, fill=white!20, text width=6em, 
	text centered, minimum height=15em]
\tikzstyle{binderdriver} = [draw, fill=white!20, text width=20em, 
	text centered, minimum height=3em, minimum width=24em]
\tikzstyle{stub} = [draw, dashed, rotate=90, minimum height=1.4em, 
	text centered]
\tikzstyle{activity} = [draw, fill=white!20,
	minimum height=13.3em,  minimum width=5em]
\tikzstyle{proxy} = [draw, dashed, rotate=90, minimum height=1.2em, 
	minimum width=1em, text centered]
\def\blockdist{5em}
\def\edgedist{3em}
    \draw node[draw, text width=8cm, color=red, text centered] (binder) at (0,0) {Binder};
    
    \draw node[draw, text width=4.6cm, text height=1.1cm] (apps) at (-1.1,2.79) {};   \draw node[draw, text width=4.4cm] (androidApp) at (-1.1,2.5) {\scriptsize Activity|Service|Provider|Receiver};
    
    \draw node[draw] (serviceManager) at (2.6,4) {Service Manager};
  \draw node[draw, below=.5cm of serviceManager.south] (systemServer) {System Server};
  
    \draw node[text width=3.1cm, above =.2em of androidApp, text centered] (nativeApp) {\small Android Application}; 
    \tikzset{communicationMean/.append style={draw, dotted, minimum height=2em}}
  \draw node[communicationMean, text width=2cm, text centered] (intent) at (-3,1) {Intent};   \draw node[communicationMean, text width=2cm, text centered] (queryUri) at (0,1) {Query};   \draw node[communicationMean, text width=2cm, text centered] (proxyStub) at (2.8,1) {Proxy/Stub};

    \path [draw, -, dashed] (intent) -- (intent |- binder.north);   \path [draw, -, dashed] (proxyStub) -- (proxyStub |- binder.north);   \path [draw, -, dashed] (queryUri) -- (queryUri |- binder.north);   
    \path [draw, -, dashed, transform canvas={xshift=-1.9cm}] (apps) -- (apps |- intent.north);
  \path [draw, latex-latex] ([yshift= -.4 cm]apps) -| ([xshift=-.2cm]proxyStub.north);
  \path [draw, -, dashed, transform canvas={xshift=1.1cm}] (apps) -- (apps |- queryUri.north);
  
    \path [draw, latex-latex] (apps) |- (serviceManager);
  \path [draw, latex-latex, transform canvas={xshift=.5cm}] (systemServer) -- (systemServer |- proxyStub.north); 
  
\end{tikzpicture}
 }
\caption{Android Communication Overview.}
\label{fig:android-framework}
\end{center}
\end{figure}
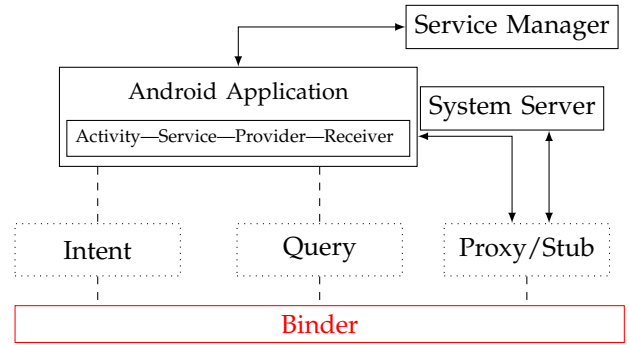

Components communicate with one another through the binder, the Android-specific Inter Process Communication (IPC) mechanism, and Remote Method Invocation (RMI) system. 
Components do not communicate with the binder directly but instead rely on three high-level abstractions of communication called \emph{intent}, \emph{query} and \emph{proxy}.
Figure \ref{fig:android-framework} focuses on those communications at the Java level of the Android framework.
It shows that an application communicate with the system server (and thus system services) through proxies and stubs (abstraction on top of the binder).
\textbf{Intent.}
Intents describe operations to be performed. 
They are used to start a new user interface screen (Activity), trigger a component which listens to intents (BroadcastReceiver) or communicate with services.
\textbf{Query/Uri.}
Queries are used to communicate with content provider components (which share data for instance through a database). 
Queries use Uniform Resource Identifier (URI) to indicate the target provider component on which the query must be performed. 
\textbf{Proxy/Stub.}
System services extend stub classes which describe methods they must implement. 
System services are mainly used by application to access system resources.
They are accessed by other components through their public interface called proxy.
System services are running in the system server and are registered to the service manager. 
An application can get a reference to a registered service through the service manager and can then communicate with the service through its proxy (which uses the binder). 
\makeatletter{}\begin{table}
  \scriptsize
 \begin{center}
\resizebox{\columnwidth}{!}{
  \begin{tabular}{|cl|}
\hline
    \method{int} & \method{checkPermission  (String, int, int)} \\ \hline
    \method{int} & \method{checkCallingPermission  (String)} \\ \hline
    \method{int} & \method{checkCallingOrSelfPermission  (String)} \\ \hline
    \method{void} & \method{enforcePermission  (String, int, int, String)} \\ \hline
    \method{void} & \method{enforceCallingPermission  (String, String)} \\ \hline
    \method{void} & \method{enforceCallingOrSelfPermission  (String, String)} \\ \hline
  \end{tabular}
}
\caption{List of Permission Check Methods of the \class{an\-droid.con\-tent.Con\-text} Class (since Android 1.0 / API Level 1)}
\label{table:permissionCheckMethods}
 \end{center}
\end{table}
 
\section{Static Analyses for Android}
\label{sec:application-on-android}
	\makeatletter{}
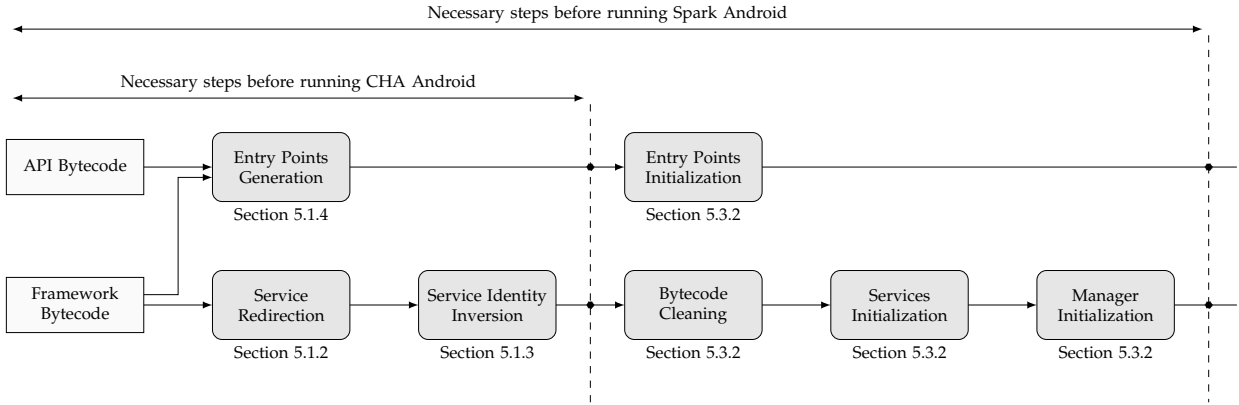
\begin{figure*}
\begin{center}
\resizebox{\textwidth}{!}{
\makeatletter{}\tikzstyle{nodeState} = [draw, 
  text centered, 
  text width=20mm, 
  minimum height=10mm, 
  inner sep = 0mm, 
  outer sep = 0mm,
  rounded corners, 
  fill=black!10]
\tikzstyle{nodeData} = [draw, 
  text centered, 
  text width=20mm, 
  minimum height=8mm, 
  inner sep = 0mm, 
  outer sep = 0mm, 
  fill=black!2]
\tikzstyle{nodeMethod} = [draw, text centered, text width=2.7cm]
\tikzstyle{nodeEvent} = [text centered, text width=25mm, inner sep=.1cm, font=\small, densely dotted, draw]
\tikzstyle{dotShape} = [circle, text centered, text width=0mm, inner sep=.01cm, font=\small, densely dotted, draw, fill=black, minimum height=1mm]
\begin{tikzpicture} 
\scriptsize
  \path 
    (0,0) node[nodeData] (apiBytecode) {API Bytecode}
    +(-1, 1) node[] (chaLeft) {}
    +(-1, 2) node[] (sparkLeft) {}
  +(1.5,-1) node[inner sep = 0cm, outer sep =0cm, minimum size=0cm] (nothing005) {}
  ++(3,0) node[nodeState,  label=below:Section~\ref{sub:entryPointsConstructionCHA}] (entryPointGeneration) {Entry Points \\ Generation}
  +(4.5,0) node[dotShape] {}
  +(4.5, 1) node[] (chaTop) {}
  +(4.5, -3.5) node[] (chaBottom) {}
  ++(6,0) node[nodeState,  label=below:Section~\ref{sub:entryPointsConstructionSpark}] (entryPointInitialization) {Entry Points \\ Initialization}
  ++(8,0) node[] (nothing003) {}
  ++(-17,0) node[] (nothing001) {}
  ++(0,-2) node[nodeData] (frameworkBytecode) {Framework Bytecode}
  ++(3,0) node[nodeState,  label=below:Section~\ref{sec:binder}] (serviceRedirection) {Service Redirection }
  +(4.5,0) node[dotShape] {}
  ++(3,0) node[nodeState,  label=below:Section~\ref{sec:binder-identity}] (serviceIdentityInversion) {Service Identity Inversion}
  ++(3,0) node[nodeState,  label=below:Section~\ref{sub:processingTime}] (bytecodeCleaning) {Bytecode Cleaning}
  ++(3,0) node[nodeState,  label=below:Section~\ref{sec:servicesInit}] (serviceInit) {Services Initialization}
  ++(3,0) node[nodeState,  label=below:Section~\ref{sec:managersInit}] (managerInit) {Manager Initialization}
  +(1.5,0) node[dotShape] {}
  +(1.5,2) node[dotShape] {}
  +(1.5,4) node[] (sparkTop) {}
  +(1.5,-1.5) node[] (sparkBottom) {}
  ++(2,0) node[] (nothing002) {}
  ;
  \path [draw, -latex] (apiBytecode) -- (entryPointGeneration);
  \path [draw, -latex] ([yshift=1.5 mm]frameworkBytecode.east) -| (nothing005) |- ([yshift=-1.5 mm]entryPointGeneration.west);
  \path [draw, -latex] (frameworkBytecode) -- (serviceRedirection);
  \path [draw, -latex] (serviceRedirection) -- (serviceIdentityInversion);
  \path [draw, -latex] (serviceIdentityInversion) -- (bytecodeCleaning);  
  \path [draw, -latex] (bytecodeCleaning) -- (serviceInit);
  \path [draw, -latex] (serviceInit) -- (managerInit);
  \path [draw, -] (entryPointInitialization) -- (nothing003);
  \path [draw, -] (managerInit) -- (nothing002);
  \path [draw, -latex] (entryPointGeneration) -- (entryPointInitialization);
  \path [draw, dashed, -] (chaTop) -- (chaBottom);
  \path [draw, dashed, -] (sparkTop) -- (sparkBottom);
  \path [draw, latex-latex] (chaLeft) -- (chaTop) node [pos=0.5, above] {Necessary steps before running CHA Android};
  \path [draw, latex-latex] (sparkLeft) -- (sparkTop) node [pos=0.5, above] {Necessary steps before running Spark Android};;
\end{tikzpicture}
 
}
\end{center}
\caption{Bytecode Processing Before \chaIntelligent{}/\sparkIntelligent{} Analyses. 
Entry points are generated using methods from the Android SDK API bytecode.
Bytecode from the framework is transformed to redirect call to services to actual service classes, bypassing the ICC glue code.
\chaIntelligent{} requires entry point generation, service redirection and service identity.
\sparkIntelligent{} is more precise thus requires proper entry points, services, and managers initialization.
}
\label{fig:bytecode-processing}
\end{figure*}
Our goal is to define static analyses for extracting permission checks.
In essence, each analysis constructs a call graph from the bytecode, finds permission check methods and extracts permission names.
Obtaining a meaningful call graph is challenging. We ran the default Soot's CHA-Naive (Class Hierarchy Analysis) on Android 4.0.1. 
It takes more than one week and outputs $31,458/50,029$ (64\%) methods with no permissions, one method with a single permission\footnote{
This is the \perm{INTERNET} permission checked in class \class{android.webkit.WebSettings.}} and $18,381/50,029$ (36\%) entry points (methods) that each needs more than 100 high-level permissions. 
This is not meaningful. 
The reason is that Android has been implemented using the object-oriented paradigm and there are many subclasses of the core classes (e.g., of Service\footnote{\url{https://developer.android.com/reference/android/app/Service.html}} , Activity\footnote{\url{https://developer.android.com/reference/android/app/Activity.html}}, etc.).
By construction, CHA outputs that all clients of those classes call all their subclasses. 
This results in an explosion of edges in the call graph and consequently an explosion of required permissions.
\emph{The main challenge for defining static analyses for extracting permission checks is to get a precise call graph.}
We still aim at using CHA, but we need to customize it for Android. 
We also aim at using Soot's Spark \cite{lhothend03spark}, an Andersen-like points-to analysis.
Our motivations for running CHA are as follows. 
First, it enables us to identify key Android-specific analysis components.
Those components can be reused with benefits in more sophisticated analyses such as Spark.
Second, it gives us a baseline for assessing the improvements given by Spark.
Third, it gives a list of API methods with no permission which do not require to be analyzed by Spark.
Eventually, the best-of-breed of Android specific analysis components and Spark enable us to obtain a fairly precise permission map. 
Figure \ref{fig:bytecode-processing} represents Android-specific components that manipulate the framework bytecode, and generate and initialize entry points. \chaIntelligent{}, the customized version of CHA for Android, requires 
generation of the entry point, presented in Section \ref{sub:entryPointsConstructionCHA}, 
service redirection, described in Section \ref{sec:binder}, and 
service identity inversion, detailed in Section \ref{sec:binder-identity}.
In addition to those components, \sparkIntelligent{}, the customized version of Spark for Android, requires proper entry point initialization as well as services and managers initialization. Those components are described in Section \ref{sec:servicesInit}.
In our experiments, the call graphs are generated from the 50,029 entry points found in the Android API version \androidVersion{}.
All the analyses use Soot \cite{cetus11soot}, a widely used framework for the static analysis of Java programs.
The experiments run on a Intel(R) Xeon(R) CPU E5620  @ 2.40\,GHz running GNU/Linux Debian 3.11; 
the Java virtual machine 1.7.0 is given 4\,Gb of heap memory.
The Android version used in the experiments is 4.0.1 unless otherwise specified.
Section \ref{sub:common-components} presents the different components to modify the bytecode and to extract permissions from the call graph.
Section \ref{sub:intelligentCHA} describes the \chaIntelligent{} analysis and Section \ref{sub:intelligentSpark} the \sparkIntelligent{} analysis.
\makeatletter{}
\subsection{Common Components for CHA and Spark}
\label{sub:common-components}
In this section we present three techniques that are required for both CHA and Spark.
String analysis is used to extract the permission names from the call graph.
Service redirection enables the call graph construction algorithm to link the service caller to the service itself by bypassing the ICC glue code.
Finally, service identity inversion removes code from the call graph which is executed as a system service itself and thus is not relevant from the entry point caller's point of view.
\makeatletter{}\subsubsection{String Analysis for Extracting Permissions from Permission Enforcement Points}
\label{sec:stringAnalysis}
A basic call graph can only give the number of permission checks but not the actual names of the checked permissions because of the lack of string analysis to extract permission names from the bytecode.
As explained in Section \ref{subsec:service}, Permission Enforcement Points (PEPs) are method calls to \PEPnbr{}
methods of classes \texttt{Con\-text} and \texttt{Con\-text\-Wrapper} 
(see Table \ref{table:permissionCheckMethods} for a list of PEPs).
Those method calls can be resolved statically. However, the actual permission(s)
that are checked are dynamically set by a String parameter or sometimes, an
array of strings. 
Thus, when a check permission method is found in the call graph, a basic
analysis is only able to tell that a permission check occurs, but not which
precise permission is checked because a call graph does not handle literal and variable resolution by itself.
To overcome this issue, we have implemented a String analysis as a Soot plugin whose pseudo code is shown in Algorithm \ref{algo:stringAnalysis}. 
Once PEPs are found, it extracts the corresponding permission(s) (line 5).
This plugin performs an intra-method analysis and manages the following scenarios:
either (1) the permission is directly given as a literal parameter, 
or (2) the permission value is initialized in a variable which is given as a 
parameter, or (3) an array is initialized with several permissions and is given
as a parameter. In every case we do a backward analysis of the method's bytecode
using Soot's unit graphs which describe relations among statements of a
method. In the case where only a single permission is given to the method, 
statements in the unit graph containing a reference to a valid Android permission
String are extracted and the permissions added to the list of the permissions
needed by the method under analysis. In case of an array, all permissions of
references to Android permission Strings are added to the list.
It can happen that the permission string cannot be found in the current method
\texttt{M$_i$}'s body.
This happens when it is referenced from a local variable initialized
with one of the current method's parameter \texttt{P}.
The solution is for the analysis to go one method down in the method call-stack (lines 6-10). 
At this point the analysis goes through the statements of \texttt{M$_{i-1}$} looking for a call to
\texttt{M}. When a call is found the parameter \texttt{P} is extracted and the
string analysis starts again from there.
\begin{algorithm}[t]
 \KwIn{Method Call Stack, Target Method, Target Method Parameter}
 \KwResult{Set of Permission Strings}
 stack $\leftarrow$ Method Call Stack\;
 tm $\leftarrow$ Target Method\;
 tp $\leftarrow$ Target Parameter\;
 pSet $\leftarrow$ set ()\;
 
 pSet $\leftarrow$ findPermission (tm, tp)\;
 \If{pSet is empty}{
  tp $\leftarrow$ getCurrentMethodParameter ()\;
  N $\leftarrow$ $size (stack) - 1$\;
  r $\leftarrow$ StringAnalysis ($stack[1...N]$, $stack[N]$, $tp$)\;
  pSet $\leftarrow$ pSet $\cup$ r\;
 }
 return pSet\;
 \caption{Concrete Permissions Names Extraction (String Analysis).\label{algo:stringAnalysis}}
\end{algorithm}
 
At this point, we have a  component to extract permission strings from the call graph.
In the next section, we present how to handle service redirection to avoid having imprecise permission sets.
\makeatletter{}\subsubsection{Service Redirection: Handling Binder-based Communication}
\label{sec:binder}
\begin{figure}[t]
\begin{center}
\makeatletter{}\tikzstyle{nodeState} = [draw, 
  text centered, 
  text width=20mm, 
  minimum height=10mm, 
  inner sep = 0mm, 
  outer sep = 0mm,
  rounded corners, 
  fill=black!10]
\tikzstyle{nodeData} = [draw, 
  circle,
  minimum height=2mm, 
  inner sep = 0mm, 
  outer sep = 0mm, 
  fill=black!2]
\tikzstyle{txt} = [
text centered, 
text width=2cm]
\tikzstyle{nodeMethod} = [draw, text centered, text width=2.7cm]
\tikzstyle{nodeEvent} = [text centered, text width=25mm, inner sep=.1cm, font=\small, densely dotted, draw]
\tikzstyle{dotShape} = [circle, text centered, text width=0mm, inner sep=.01cm, font=\small, densely dotted, draw, fill=black, minimum height=1mm]
\begin{tikzpicture} 
\scriptsize
  \path 
  (0, -5cm) node[txt] {API \\ methods}
  ++(2cm, 0) node[txt] {Binder \\ \method{transact} \\ method}
  ++(2cm, 0) node[txt] {Services \\ \method{onTransact} \\ methods}
  ++(2cm, 0) node[txt] {Services \\ target \\ methods}
  ;
  \path 
  (0,0) node[] (n0) {}
  ++(0,-6mm) node[] (n1) {}
    ++(0,-6mm) node[] (n3) {}
    ++(0,-6mm) node[nodeData, label=left:Api$_{S1.1}$] (n5) {}
    ++(0,-6mm) node[] (n7) {}
    ++(0,-6mm) node[] (n9) {}
    ++(0,-3mm) node[] (n11) {}
  ++(0,-3mm) node[]  {}
  ++(2cm, 18mm) node[nodeData] (transact) {}
  ++(2cm, 15mm) node[nodeData, label=above:S$_1$] (m0) {}
  ++(0, -3mm) node[nodeData] (m1) {}
  ++(0, -3mm) node[nodeData] (m2) {}
  ++(0, -3mm) node[nodeData] (m3) {}
  ++(0, -3mm) node[nodeData] (m4) {}
  ++(0, -3mm) node[nodeData] (m5) {}
  ++(0, -3mm) node[nodeData] (m6) {}
  ++(0, -3mm) node[nodeData, label=right:S$_g$] (m7) {}
  ++(0, -3mm) node[nodeData, label=right:S$_h$] (m8) {}
  ++(0, -3mm) node[nodeData, label=right:S$_i$] (m9) {}
  ++(0, -3mm) node[] (m10) {\vdots{}}
  ++(2cm, 45mm) node[nodeData, label=right:S$_1$m$_1$ $p_0$] (s00) {}
  ++(0, -2mm) node[nodeData, label=right:S$_1$m$_2$ $p_0$] (s01) {}
  ++(0, -2mm) node[nodeData, label=right:S$_1$m$_3$ $p_1$] (s02) {}
  ++(0, -2mm) node[nodeData, label=right:S$_1$m$_4$ $-$] (s03) {}
  ++(0, -2mm) node[nodeData, label=right:S$_1$m$_5$ $p_2$] (s04) {}
  ++(0, -2mm) node[nodeData, label=right:S$_1$m$_6$ $p_0$] (s05) {}
  ++(0, -6mm) node[nodeData, label=right:S$_2$m$_1$ $p_3$] (s10) {}
  ++(0, -2mm) node[nodeData, label=right:\vdots{}] (s11) {}
  ++(0, -2mm) node[nodeData] (s12) {}
  ++(0, -2mm) node[nodeData] (s13) {}
  ++(0, -2mm) node[nodeData] (s14) {}
  ++(0, -2mm) node[nodeData] (s15) {}
  ++(0, -6mm) node[nodeData, label=right:S$_3$m$_1$ $p_6$] (s20) {}
  ++(0, -2mm) node[nodeData, label=right:\vdots{}] (s21) {}
  ++(0, -2mm) node[nodeData] (s22) {}
  ++(0, -2mm) node[nodeData] (s23) {}
  ++(0, -2mm) node[nodeData] (s24) {}
  ++(0, -2mm) node[nodeData] (s25) {}
  ++(0, -3mm) node[] {\vdots{}}
  ;
    \path [draw, dashed, -latex, black] (n5)  edge[bend left=25] node {} (s00);
        \path [draw, -latex] (n5) -- (transact);
      
  \path [draw, -latex] (transact) -- (m0);
  \path [draw, -latex] (transact) -- (m1);
  \path [draw, -latex] (transact) -- (m2);
  \path [draw, -latex] (transact) -- (m3);
  \path [draw, -latex] (transact) -- (m4);
  \path [draw, -latex] (transact) -- (m5);
  \path [draw, -latex] (transact) -- (m6);
  \path [draw, -latex] (transact) -- (m7);
  \path [draw, -latex] (transact) -- (m8);
  \path [draw, -latex] (transact) -- (m9);
  
  \path [draw, -latex] (m0) -- (s00);
  \path [draw, -latex] (m0) -- (s01);
  \path [draw, -latex] (m0) -- (s02);
  \path [draw, -latex] (m0) -- (s03);
  \path [draw, -latex] (m0) -- (s04);
  \path [draw, -latex] (m0) -- (s05);
  \path [draw, -latex] (m1) -- (s10);
  \path [draw, -latex] (m1) -- (s11);
  \path [draw, -latex] (m1) -- (s12);
  \path [draw, -latex] (m1) -- (s13);
  \path [draw, -latex] (m1) -- (s14);
  \path [draw, -latex] (m1) -- (s15);
  \path [draw, -latex] (m2) -- (s20);
  \path [draw, -latex] (m2) -- (s21);
  \path [draw, -latex] (m2) -- (s22);
  \path [draw, -latex] (m2) -- (s23);
  \path [draw, -latex] (m2) -- (s24);
  \path [draw, -latex] (m2) -- (s25);
\end{tikzpicture}
 
\end{center}
\caption{
The number of edges explodes when an API method reaches the \method{transact} method of the \class{Binder} class.
This node leads to an explosion in the number of permission since it reaches all services' \method{onTransact} methods and each of those reaches all methods of their service. Those methods check for different permissions.
Solving this problem boils down to short-circuit the low level \method{transact} and \method{onTransact} methods to directly reach the method of interest.
The solution is represented by the dashed arrow which directly links an API method to its corresponding method in the right service. 
Thus, the API method is not mapped to permissions $\{p_0, p_1, p_2, p_3, p_6, ... \}$ but only to permission $p_0$.
}
\label{fig:permission-size-explosion}
\end{figure}
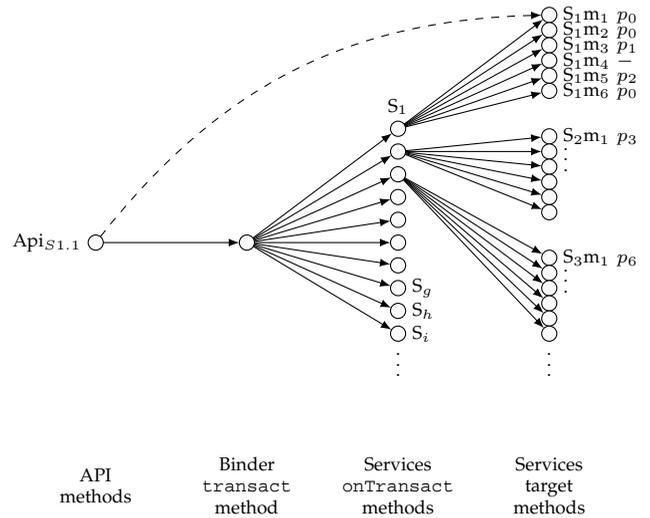
\textbf{Permission Size Explosion.} 
A call to a service method usually goes through a \texttt{manager} which gets a reference to a system service called \texttt{proxy}.
It is always a method call on a \texttt{proxy} which results in data marshaling from the proxy through the \texttt{binder} to the \texttt{stub} on top of which lays the real system service method.
All data transfers between the proxy and stub go through the \method{transact()} method which calls the \method{onTransact()} method. 
This method calls the right method on the system service side according to an integer value.
This integer value is not determined when doing a static analysis.
Thus, as illustrated in Figure \ref{fig:permission-size-explosion}, all methods of system services are added as edges in the call graph.
Moreover, as all system services implement a \texttt{stub}, when constructing the call graph using CHA, all system services stubs' \method{onTransact()} methods are potential method calls from every method call on a proxy object and are thus added to the graph.
A consequence of this is the explosion of the permission set size we observe when running CHA.
In short, when doing a naive analysis from the point of view of services, any system service method call does have edges to all methods of every system service.
\textbf{Service Redirection.}
Figure \ref{fig:android-system-service} illustrates a communication between an application and a service.
The communication is done through the binder.
As explained in the previous paragraph, the problem is that analyzing binder based communications leads to an explosion in the number of permission.
The solution, illustrated Figure \ref{fig:permission-size-explosion}, is to bypass the binder (proxy/stub) mechanism by directly connecting a call to a service method to the corresponding method within the remote service.
In Figure \ref{fig:android-system-service} edges from method \method{r.getPassword()} to the binder and from the binder to service method \method{getPassword()} are removed. 
Only the direct edge from the calling method to the called method (not shown in the Figure) is kept.
As presented in Figure \ref{fig:bytecode-processing} this is the first transformation done on the bytecode of the Android framework.
 
We now know how to redirect system services properly. 
However, it may happen that system services execute code on their behalf and not on the behalf of the original caller.
The next section explains how we remove this code from the call graph.
\makeatletter{}\subsubsection{Service Identity Inversion}
\label{sec:binder-identity}
In Android, services can call other services either with the identity of the initial caller (by default) or with the identity of the service itself. In the later case, remote calls are within \method{clearIdentity()} and \method{restoreIdentity()} method calls. 
When using the service's own identity, permission checks are not done against the caller's declared permissions, but against the service's declared permissions. 
Since our goal is to compute the permission gap of an application (and not of system services), we can safely discard all permission checks that occur between calls to \method{clearIdentity()} and \method{restoreIdentity()}.
For instance, let us assume that service S requires and declares permission $\theta$ which is not declared by application A. 
If A calls S, the code of S is executed  with the identity of A itself which would require A to declare $\theta$. 
To avoid this, the portion of code requiring $\theta$ is executed with S's identity. 
When we encounter calls to \method{clearIdentity()} or \method{restoreIdentity()}, we use an intra-procedural flow-sensitive analysis to discard permission checks that occur between those calls. 
Figure \ref{fig:bytecode-processing} shows that the Service Identity Inversion step is done after the Service Redirection transformation.
 
Sections \ref{sec:binder} and \ref{sec:binder-identity} explain how to construct a call graph taking into account specificities of the Android system.
As we do analyze a framework and not a traditonnal application, the call graph construction starts from entry points of the framework and not from a main method.
The next section explains how we construct a call graph from entry points.
\makeatletter{}
\subsubsection{Entry Points Handling for CHA}
\label{sub:entryPointsConstructionCHA}
In the case of an API (such as the Android API), the problem is that there is no ``main'' but $N$ classes totalizing $M$ entry point methods. 
Our solution is to build one call graph per public method of the Android API by creating one fake method \texttt{m$_{class_i}$} ($i \in (1,\dots,N)$) per public class of the framework (for Android, \texttt{android.*} and \texttt{com.android.*}). 
The role of method \texttt{m$_{class_i}$} is to create an instance \texttt{o} of \texttt{$class_i$} and to call all methods of \texttt{$class_i$} on \texttt{o}.
We also build a unique artificial main calling all \texttt{m$_{class_i}$} methods.
This main method is the unique start point of the analysis.
As presented in Figure \ref{fig:bytecode-processing}, entry points are constructed using methods from the Android API.
 
Section \ref{sub:intelligentCHA} presents \chaIntelligent{} which leverages the service redirection, service identity inversion and entry point construction components.
 
\makeatletter{}
\subsection{\chaIntelligent{}}
\label{sub:intelligentCHA}
We perform the map construction with CHA for three reasons.
First, it enables us to identify key Android-specific analysis components that can be reused with benefits in more sophisticated analyses such as Spark.
Then, it gives us a baseline for assessing the improvements given by Spark.
Finally, it gives a list of more than 30k API methods with no permission which do not require to be analyzed by Spark.
\chaIntelligent{} is a CHA-based static analysis for extracting permission checks on the Android framework.
It uses the string analysis presented in Section \ref{sec:stringAnalysis},
the service redirection (Binder) of Section \ref{sec:binder},
and the service identity inversion explained in Section \ref{sec:binder-identity}. We enrich it with an optimization that we now describe.
\makeatletter{}\subsubsection{Call Graph Search Optimization}
\label{sec:graphOptimization}
Section \ref{sec:stringAnalysis} describes how to extract permission names.
This Section explains how permission names are propagated through the graph from PEPs.
Algorithm \ref{algo:permissionsExtraction} propagates permission sets through the graph.
It proceeds in three steps.
The first step (line 2) traverses the graph using depth first search and keeps track of the methods already visited.
During the traversal it finds where permissions are checked and extracts the permission names (see string analysis above). 
This first step makes the analysis much faster than the naive approach since no method is analyzed more than once.
Steps two and three make sure that permissions of already analyzed method are propagated in the graph.
During the second step (lines 3-4) we use Tarjan's algorithm \cite{tarjan:connectedComponents} to replace Strongly Connected Components (SCC) from the graph by a single node.
This essentially removes loops from the graph and simplifies the propagation of permission names.
During this step one has to be careful not to remove essential parts of the graph such as methods that check permissions since permissions are not propagated at this stage.
Concretely, if a check permission method is part of an SCC it must not be removed from it otherwise permissions mapped to this method would not be propagated and thus be lost.
The third and last step (line 5) propagates permissions throughout the graph.
This algorithm has a linear complexity in the number of nodes and edges. 
During the first step the graph is searched using depth-first search and methods are never analyzed twice: this step is bound linear in the number of edges and nodes.
Tarjan's algorithm is bound linear in the number of nodes and edges. 
The last step propagates permissions through a depth first search of the graph where SCCs are replaced.
\begin{algorithm}[t]
 \caption{Permissions Extraction and Propagation.\label{algo:permissionsExtraction}}
 \KwIn{Call Graph}
 \KwResult{Set of Methods with their Permission Sets}
 g1 $\leftarrow$ Call Graph\;
 DephtFirstSearchAndPermissionExtraction (g1)\;
 SCC $\leftarrow$ TarjanFindSCC (g1)\;
 g2 $\leftarrow$ ReplaceSCC (g1, SCC)\;
 PropagatePermissions (g2)\;
\end{algorithm}
 
\subsubsection{Empirical Results}
\textbf{Permission Strings Resolution.}
Let us now analyze the efficiency of the string analysis.
The distribution of the results of string analysis is presented in Table~\ref{table:permDistribution}. 
We observe that 91.89\% of the permission string analyses only check a single
permission and that 83.25\% of the analysis the permission string can directly
be determined as a literal parameter.
Hence, it is a common practice in the Java codebase of Android to (1) protect a method with only one or two permissions and (2)
to make reference to permission strings and call the check permission method in the same method body.
Those results show that for \emph{99.08\%} of permission checks the permission
string is found using a string analysis.
Sometimes (0.92\%), it is not possible to resolve permission strings: in 12
cases permissions are related to URIs; in two cases permissions are read from the Binder (Parcel).
\begin{table}
\begin{center}
\begin{tabular}{p{5cm}|lr}
 Total \# analyses       & 1,516 (100.00\%)\\ \hline \hline
 \multicolumn{1}{l}{}String found  &     	\\ \hline \hline
 total & 1,502 (99.08\%) \\ \hline
 with 1 permissions      & 1,393 (91.89\%) \\ 
 with 2 permissions      & 109  (7.19\%) \\ \hline 
 with only direct strings& 1,262 (83.25\%) \\
 with flow analysis      & 183  (12.07\%)\\
 with strings in array   & 57 (3.76\%)\\ \hline \hline
 \multicolumn{1}{l}{}String not found  &     	\\ \hline \hline
 total & 14 (0.92\%) \\ \hline
  with URI read perm.     & 6 (0.40\%)\\
 with URI write perm.    & 6 (0.40\%)\\ 
 with read from parcel   & 2 (0.13\%)\\ \hline 
\end{tabular}
\end{center}
\caption{\label{table:permDistribution}The Kinds of Permission Specification as Found by Our String Extraction Analysis.}
\end{table}
\textbf{Execution time.}
On Android, \chaIntelligent{} analyzes 50,029 entry points  in 4 minutes user time or 10 minutes real time.
This shows that \chaIntelligent{} is able to scale on a large scale real world Framework.
\begin{table}[h]
\begin{center}
\begin{tabular}{p{5cm}|r}
  Permission Set     & \# entry points \\ \hline \hline
with 0 permissions   & 32,924 (65.8\%) \\
with 1 permissions   & 39 (0.08\%) \\
with 2 permissions   & 55 (0.12\%) \\
with \textgreater~65 permissions & 17,011 (34.0\%)\\ \hline
                     & 50,029 (100\%) \\
\end{tabular}
\end{center}
\caption{\label{table:chaIntelligentPSets}\chaIntelligent{} Permission Sets.}
\end{table}
\textbf{Entry Point Permission Sets.}
Running \chaIntelligent{} yields Table \ref{table:chaIntelligentPSets} which shows the permission set size for the entry points. 
As \chaIntelligent{} correctly models system service communications, the number of entry points requiring no permissions increases from 64\% to 65.1\% (31,458 to 32,429) (some service methods are not protected by permissions). 
The number of entry points with one and two permissions increases from less than 0.01\% to 0.08\% (1 to 39) and from 0\% to 0.12\% (0 to 55) respectively (service method redirection avoids explosion in the number of edges in the call graph and thus the number of permissions).
Nevertheless, 34\% (17,011) of entry points still have an over-approximated permission set.
This is caused by the imprecision of the points-to set of CHA. 
This results in an explosion in the number of permissions.
An improvement would be to develop other domain specific optimizations:
handling other Android specific points (e.g. content providers, handlers and messages) is similar to handling service communications and would not have an impact on the contributions of this paper.
The following Section \ref{sub:intelligentSpark} presents the Spark based analysis.
The analysis tackles Spark specific issues such as entry point initialization or Android specific issues such as service initialization.
 
\makeatletter{}
\subsection{\sparkIntelligent{}}
\label{sub:intelligentSpark}
We run Spark in context\--insensitive, path\--insensitive, flow\--insensitive, field\--sensitive mode  to generate the call graph. 
In context-insensitive mode, every call to the same method is merged to a single edge independently of the context (receiver and parameters values).
A path-insensitive analysis ignores conditional branching hence takes into account all paths of method bodies.
The call graph construction is flow-insensitive since it does not consider the order of executions of instructions.
It is also field-sensitive because it differentiates the points-to solution associated with different named object fields.
We first run a naive version of \sparkIntelligent{} in Section \ref{sub:naiveNaiveSpark} to illustrate the need to correctly initializing objects on which API methods are called as well as method's parameters. 
Section \ref{sub:sparkSpecificAnalysisComponents} describes how we initialize entry points.
It also explain another Spark subtlety: why and how system services must be initialized.
\subsubsection{Naive Usage of Spark}
\label{sub:naiveNaiveSpark}
As for CHA, we ``naively'' run off-the-shelf Spark to get a first understanding of the main problems that occur when analyzing the Android API.
This gives us a key insight, Spark discards 96\% of the API methods to be analyzed.
The reason is that Spark does not work on receiver objects whose value is \emph{null} 
(i.e. methods called on references initialized by default with null do not appear in the graph).
The four percents of analyzed methods are Java static methods which can be called without instantiating their classes.
This means it is not possible to run a Spark based analysis without correctly initializing entry points.
Even with key Android-specific static analyses of CHA, a naive usage of Spark completely fails.
Consequently, we need Spark specific analysis components.
\subsubsection{Spark Specific Analysis Components}
\label{sub:sparkSpecificAnalysisComponents}
\textbf{Processing Time.}
\label{sub:processingTime}
Our first experiments show that Spark does not scale to the size of the Android framework.
As we experience that Spark is time consuming when processing some entry points, we empty specific methods of certain classes to be able to compute permissions sets in a realistic amount of time (i.e., less than one day).
Analyzing time consuming entry points always leads to the windowing system classes.
The windowing system is at the heart of Android components such as activities. 
It is responsible for the GUI (Graphical User Interface) management, and has relationships with numerous GUI abstractions such buttons or text fields and methods to start Android components such as other activities.
When the call graphs hits a component of the windowing system it can grow in such huge proportion, because of the imprecision in the points-to sets, that the search in it triggers a timeout.
We make the hypothesis that classes responsible for GUI rendering and the windowing system management do not link to any permission check.
Thus, we remove code of their methods and launch the experiments again. 
Removing the code means that  (1) Spark does not construct the call graph for this code and thus that (2) the traversal of the call graph is much faster.
With those modifications, the computation time of the permission map is much faster, terminates in less than 11 hours and does not trigger any timeout.
\textbf{Entry Points Handling for Spark.}
\label{sub:entryPointsConstructionSpark}
\sparkEssential{} leverages artificial methods generated for CHA (see Section \ref{sub:entryPointsConstructionCHA}).
However, it must initialize parameters of the 50,029 entry point methods of the Android API.
Each receiver object \texttt{o} on which to call Android API methods as well as every method parameter \texttt{p} are initialized by calling \texttt{ge\-ne\-ra\-te$_o$()} and \texttt{ge\-ne\-ra\-te$_p$()}, respectively.
This tailor made method generates all possible instances of type \texttt{P} (i.e., over-approximation). 
Parameter initialization is necessary since one does not know a priori the effect of parameters on permission checks. 
Since Spark is field-sensitive, non-initialized parameters result in missing edges in the call graph.
\begin{figure}[h!]
\begin{center}
  \makeatletter{}\tikzstyle{nodeApp} = [draw, dotted, text width=.95\columnwidth, font=\scriptsize, inner sep=0cm]
\tikzstyle{nodeSystem} = [draw, text width=.95\columnwidth, font=\scriptsize, inner sep=0cm]
\begin{tikzpicture} [node distance = .1em]
\node[nodeApp] (node1)  {
\begin{lstlisting}
AccountManager m = getSystemService("account");
m.getPassword(a);
\end{lstlisting}};
\node[nodeSystem, below = .1cm of node1] (node2) {
\begin{lstlisting}[label={lst:dummyMain}]
public class AccountManager {
  IAccountManager mServ; 
  public String getPassword(Account a) {
    // the callgraph stops here because 
    // mService  is null (see Figure 4)
    return mServ.getPassword(a);
  }
}
\end{lstlisting}};
\node[draw, dotted, below = .6cm of node2, minimum width=.6cm, minimum height=.3cm, xshift = -3cm, label=right:\scriptsize Application code] (legend1) {};
\node[draw, right = 3cm of legend1, minimum width=.6cm, minimum height=.3cm, label=right:\scriptsize API/System code] (legend2) {};
\end{tikzpicture}
 
\caption{How Spark Discards Call Graph Edges Because of ''null'' Objects.}
\label{fig:sparkDiscardsEdges}
\end{center}
\end{figure}
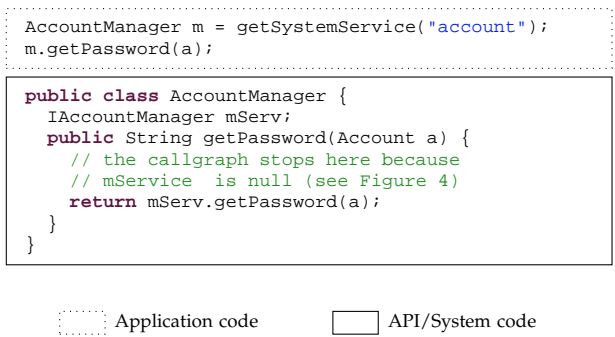
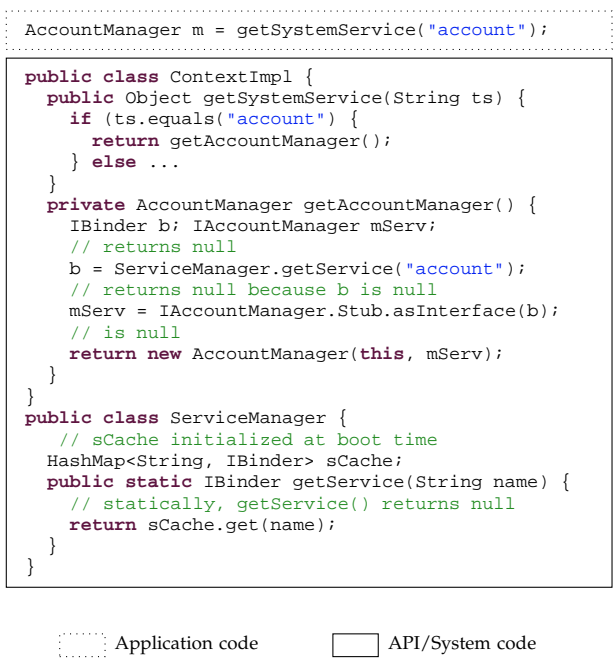
\begin{figure}[h!]
\begin{center}
	\makeatletter{}\tikzstyle{nodeApp} = [draw, dotted, text width=.95\columnwidth, font=\scriptsize, inner sep=0cm]
\tikzstyle{nodeSystem} = [draw, text width=.95\columnwidth, font=\scriptsize, inner sep=0cm]
\begin{tikzpicture} [node distance = .1em]
\node[nodeApp] (node1)  {
\begin{lstlisting}
AccountManager m = getSystemService("account");
\end{lstlisting}};
\node[nodeSystem, below =.1cm of node1] (node2) {
\begin{lstlisting}[label={lst:dummyMain}]
public class ContextImpl {
  public Object getSystemService(String ts) {
    if (ts.equals("account") {
      return getAccountManager();
    } else ...
  }
  private AccountManager getAccountManager() {
    IBinder b; IAccountManager mServ;
    // returns null
    b = ServiceManager.getService("account"); 
    // returns null because b is null
    mServ = IAccountManager.Stub.asInterface(b);
    // is null
    return new AccountManager(this, mServ); 
  }
}
public class ServiceManager {
   // sCache initialized at boot time
  HashMap<String, IBinder> sCache;
  public static IBinder getService(String name) {
    // statically, getService() returns null
    return sCache.get(name); 
  }
}
\end{lstlisting}};
\node[draw, dotted, below = .6cm of node2, minimum width=.6cm, minimum height=.3cm, xshift = -3cm, label=right:\scriptsize Application code] (legend1) {};
\node[draw, right = 3cm of legend1, minimum width=.6cm, minimum height=.3cm, label=right:\scriptsize API/System code] (legend2) {};
\end{tikzpicture}
 
\caption{How Spark Propagates "null" Due to Initialization that is not Statically Visible.}
\label{fig:sparkPropagatesNull}
\end{center}
\end{figure}
\textbf{Importance of Service Initialization for Spark.}
A Spark based approach does require proper initialization of the analyzed modules of the Android framework.
The reason is that, as presented in Figures \ref{fig:sparkDiscardsEdges} and \ref{fig:sparkPropagatesNull}, skipping the initialization phase may result in important fields, containing references to system services for instance, to only point-to \emph{null}.
Spark does not generate edges for method calls on references which can only point to \emph{null}. 
Figure \ref{fig:sparkDiscardsEdges} represents a code snippet which retrieves an \class{AccountManager} object and calls method \method{getPassword()} on it.
At this point \class{AccountManager}'s service reference \object{mServ} can only point to \emph{null}.
Thus, \method{mServ.getPassword()} cannot be executed and would not be represented in a field-sensitive call graph.
In other words, Spark generates an edge for the \class{AccountManager} object but not for the service method call within it because the service reference (\texttt{mServ}) points to \emph{null}.
 
This \class{AccountManager} object is created by the \class{Context} class as described in Figure \ref{fig:sparkPropagatesNull}.
To simplify, only \class{AccountManager} objects are created in \method{getSystemService()}.
To create an \class{AccountManager} object a reference to the AccountManagerService is required.
This reference is fetched through a call to \method{getService()}. 
However, since \class{ServiceManager} has not been initialized, \class{ServiceManager}'s sCache map is empty.
So, \method{getService()} always returns \emph{null}.  
\textbf{Service Initialization for Static Analysis}
\label{sec:servicesInit}
As detailed in Section \ref{sec:veryTechDetails}, system services are initialized in the \class{SystemServer} class.
Methods from this class are not present in the call graph generated from entry points of the Android API since they are only called at system boot time.
To simulate system services initialization we create a static object and an initialization method for each concrete system service. 
Those objects are initialized by adding edges to the service initialization methods to the call graph.
Moreover, the original bytecode is modified to replace calls to \method{getService} by a reference to the newly created static objects.
\textbf{Manager Initialization for Static Analysis}
\label{sec:managersInit}
Android applications have two possibilities to communicate with system services
\begin {itemize}
  \item The first possibility is to directly get a reference to the service\footnote{also called a \emph{binder} to the service} through the service manager and then to call remote procedures of the service
  \item The other possibility is to use another interface called \class{Manager}. The manager is created from the system \class{Context} class and has itself a reference to the service to directly communicate with it and acts as a proxy for the application (as show in Figure \ref{fig:sparkDiscardsEdges}).
\end{itemize}
Managers are wrappers to ease communication with system services.
We redirect calls to \method{getSystemService(String s)} to our own methods.
To be able to do that, we used string analysis to compute a mapping between strings given to \method{getSystemService} and the code which initializes the corresponding manager.
Each call to \method{getSystemService} is analyzed to extract the string parameter to know to which method it must be redirected. To each string corresponds one Manager and thus one method whose role is to initialize the manager.
 
We also provide our own \method{getService()} method that returns properly initialized services as presented in Section \ref{sec:servicesInit}.
All calls to the original \method{getService()} are redirected to our own methods.
Method \method{getSystemService} returns a manager whereas method \method{getService()} returns an interface to a service.
The original bytecode of the Android framework is modified to reflect services and managers initialization. 
The resulting bytecode can be analyzed by any static analysis tool and is not specific to Soot.
\subsubsection{Empirical Results}
 
\sparkIntelligent{} runs in 11 hours.
Permission set sizes for entry points when running \sparkIntelligent{} are described in Table \ref{table:sparkIntelligentPSets}.
The number of entry points with a single permission is 471.
Furthermore, 48 entry points have a permission set of two, 10 of 3 and three have more than three permissions.
The total number of entry points is less than the one for CHA since abstract classes cannot be initialized with Spark.
No method associated with those classes is represented in the set of entry point methods.
\begin{table}[h]
\begin{center}
\begin{tabular}{p{5cm}|r}
  Permission Set     & \# entry points \\ \hline \hline
with 0 permissions   &  42,895 (98.77\%)\\
with 1 permissions   &    471 (1.08\%)\\
with 2 permissions   &      48 (0.11\%)\\
with 3 permissions   &      10 (0.02\%)\\
with > 3 permissions &       3 (< 0.01\%)\\ \hline
                     &  43,427 (100\%) \\
\end{tabular}
\end{center}
\caption{\label{table:sparkIntelligentPSets}\sparkIntelligent{} Permission Sets.}
\end{table}
 
\subsection{Recapitulation}
We have presented the core technical issues we encountered while implementing our approach. 
We think that those problems may arise in other permission-based platforms than Android, and that identifying them and their solutions can be of great help for future work.
Last not but not least, those points are crucial for replication of our results.
Section \ref{sec:empirical-study} evaluates the CHA and Spark based analyses.
 
\section{Discussion}\label{sec:empirical-study}
	\makeatletter{}
In Section \ref{sec:application-on-android} we have presented three analyses: CHA-naive, CHA-Android and Spark-Android.
CHA-naive is the default analysis provided by Soot.
CHA-Android takes into account specificities of the Android system such as service redirection and system identity inversion.
Spark-Android also take into account those specificities but leverages a more precise, field-sensitive call graph construction algorithm.
How do those three analyses perform compared to others? 
What are their limitations?
This section answers those questions.
\makeatletter{}\begin{figure}[t]
\begin{tikzpicture}
\begin{axis}[
width=\columnwidth, 
xlabel= \# of permissions, 
ylabel= \# of entry point methods,
xtick = {0, 1, 2, 3},
legend style={
at={(0,0)},
anchor=north east,at={(axis description cs:.70,.69)}},
every axis legend/.append style={nodes={right}},
scaled y ticks = false,
y tick label style={/pgf/number format/fixed}
]\pgfplotstableread{./cumulative/all.dat}
\datatablecumulative
\addplot table[y = cha-naive] from \datatablecumulative;
\addlegendentry{CHA-Naive}\addplot table[y = cha-intelligent] from \datatablecumulative;
\addlegendentry{\chaIntelligent{}} 
\addplot table[y = spark-intelligent] from \datatablecumulative;
\addlegendentry{\sparkIntelligent{}} 
\end{axis}
\end{tikzpicture}
  \caption{Cumulative Plot of the Number of Methods per Permission Set Size (The higher, the better).}
  \label{fig:nbrMethodsPerPermSetSize}
\end{figure}
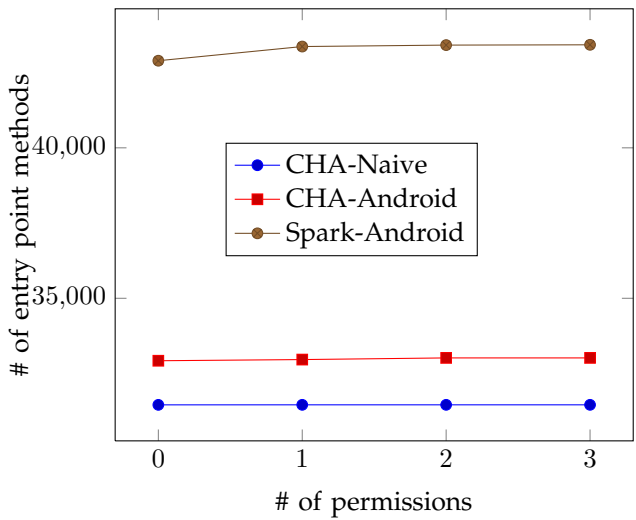
 
\subsection{CHA versus Spark}
\label{sec:chaVsSpark}
\makeatletter{}
Figure \ref{fig:nbrMethodsPerPermSetSize} is a cumulative plot of the number of entry points function of their permission set size.
By cumulative we mean that at each permission set size the number of methods is added to the number of methods at the previous permission set size.
It first shows that the more precise an analysis is, the bigger the set of entry points with zero permission will be.
This result reflects the fact that with precision, "false positive" edges are removed from the graph.
Then, the plot (\sparkIntelligent{}) highlights that, when only system services communication are handled, Spark yields the best results as it finds more methods with a permission set of one, two or three than all other analyses. 
Moreover, Spark never finds an entry point with more permission than CHA.
It finds the same permission set (with one or more permission) than with CHA for 91 entry points.
Spark finds a smaller permission set for 428 entry points.

\subsection{Comparison with PScout}
\label{sec:comparison-pscout}
\makeatletter{}
\makeatletter{}\begin{table}
\scriptsize
  \begin{center}
\begin{tabularx}{\columnwidth}{p{6cm}|X}
Permission set                          & Number of Methods\\\hline 
\#API Methods in PScout                 & 593 \\
\#API Methods in Spark and PScout       & 468 (100\%) \\ \hline 
Identical                               & 289 (61.75\%)\\ \hline 
 we find more precise permission checks & 176 (37.60\%) \\
 we find more permission checks         &   3 (0.64\%) \\
\end{tabularx}
  \caption{Comparison between Our Results (Spark-based analysis) and Pscout's ones \cite{au2012pscout} (CHA-based analysis) using Android 4.0.1.}
  \label{table:comparison-pscout}
  \end{center}
\end{table}
 
PScout \cite{au2012pscout} relies on a CHA based approach and generates a permission list for classes in the Android framework.
We only consider classes of the Android 4.0.1 API.
There are 593 methods in the results of PScout that have more than one permission and 468 methods that are both in PScout and Spark.
Among those 468 methods, 289 (61.75\%) have the same permission size in both PScout and Spark and 176 (37.60\%) have a smaller permission set size with our approach.
For instance, for method \method{exitKeyguardSecurely(\ldots)} of class \class{KeyguardManager}, PScout finds five permissions whereas Spark only one, \perm{DISABLE\_KEYGUARD}. 
The official documentation confirms that only one permission is required as well as the runtime data from Felt \cite{Felt2011a}. 
Spark also misses a permission for method \method{AudioManager.setMicrophoneMute(boolean)}.
It is because we do not handle C/C++ native code where this permission check is done.
Table \ref{table:comparison-pscout} summarizes the results of this comparison.
\emph{Our analysis yields more precise results than a pure CHA-based approach.}
Interestingly we also find three methods (0.64\%) for which our Spark approach finds more permissions than PScout's approach.
We manually checked the \class{Vibrator} class where the involved methods are defined and there is a path to a method checking permission \perm{WAKE\_LOCK}.
PScout probably did not correctly link those specific entry point methods to all methods they can reach, thus 
missing the \perm{WAKE\_LOCK} permission.
 
\subsection{Comparison with Felt et al.}
\label{sec:comparison-felt}
\makeatletter{}Let us now compare our results obtained with static analysis \cite{bartel:ase2012} with the results of Felt et al. obtained through testing ~\cite{Felt2011a}.
Both extract a list of required permissions for each method of the Android 2.2 framework.
Android 2.2 features 134 permissions, eight of them being low-level permissions that we do not analyze.
Felt et al.'s results contain 673 methods mapped to high-level permissions.
We analyze only 671 methods because 2 methods are related with application-specific objects provided in Felt's approach that are not available in our static analysis approach.
For a given method, we either find the same permission set, or a larger one.
Our method never misses a permission that Felt et al. describe.
More precisely, we infer the same permission set per method signature for 552 methods (82.3\% of commonly analyzed methods). 
There is one additional permission for 119 methods (1 additional permission for 118 methods, 2 for 1 method). There is no method for which we miss a permission,
Table \ref{table:comparison-felt} summarizes those results. Let us now discuss the discrepancy between our results.
\makeatletter{}\begin{table}
\scriptsize
  \begin{center}
\begin{tabularx}{\columnwidth}{p{6cm}|X}
Permission set                  & Number of Methods\\
\hline 
\#Methods analyzed in \cite{Felt2011a}& 1282 \\
\#Methods with HL perm. only   & 673 \\
\hline 
Identical                       &  552 (82.3\%)\\
\hline 
 we find more permission checks & 119 (17.7\%)\\
 ~~~one more                    & 118 (17.6\%)\\
 ~~~two more                    & 1 (0.1\%)\\
\hline
we find less permission checks & 0 (0\%)\\
\end{tabularx}
  \caption{Comparison between Our Results and Felt et al.'s ones \cite{Felt2011a} (Based on Testing) using Android 2.2. Only methods with high-level permissions are considered.}
  \label{table:comparison-felt}
  \end{center}
\end{table}
 
The additional permissions are due to either analyzing irrelevant code or to missing input data in Felt et al.'s approach.
In the latter case, we are able to find permissions that are checked within specific contexts that were not taken into account by the generated tests of Felt et al. 
For instance, \perm{MOUNT\-\_UNMOUNT\-\_FILESYSTEMS}  is only checked for me\-thod \method{Mount\-Service.\-shut\-down()} if the media (storage device) is ``\emph{present not mounted and shared via USB mass storage}'' (from the API documentation).
Another permission, \perm{READ\-\_PHONE\-\_STATE} is needed for me\-thod \method{Cal\-ler\-In\-fo.get\-Cal\-ler\-Id\-()} only if the phone number passed in parameter is the 
voice mail number. Those test cases were not generated by Felt's testing approach. In real applications, test generation techniques cannot guarantee a comprehensive exploration of the input space.
To us, these findings are typical when comparing a static analysis approach against a testing one: static analysis sometimes suffers from analyzing all code (including debugging and dead code, or code run in specific runtime environments), but is strong at abstracting over input data. 
On the other hand, testing must simulate as close as possible the production environment, but is cursed to always miss very specific usage scenarios.
Those results highlight the complementarity between static analysis and testing in the context of permission inference.
We think that the static analysis approach is complementary to the testing approach. 
Indeed, the testing approach yields an under-approximation which misses some permission checks whereas the static analysis approach yields an over-approximation in which those missing permission checks are found.
Using both approaches in conjunction would enable developers to obtain a lower and an upper bound of the permission gap. 
In particular, for a given Android applications, if both testing and static analysis approaches yield the same list of permissions, 
this strongly suggests that this list is the ``correct'' list of required permissions. 
As testing could miss permissions and static analysis may not model all Android specificities this cannot be a strong claim.
 
\makeatletter{}
\subsection{Soundness}
\label{sec:soundness}
We have shown in this paper that the Android framework has many specificities that may threaten the soundness of static analysis. In this context,  soundness refers to having no false negatives (no missed permission checks).
Furthermore, the concept of soundness refers to a specific scope: in our cases, checks of high-level permissions inside Android services.
For CHA and Spark-based analysis, such as PScout, \chaIntelligent{} or \sparkIntelligent{}, the manipulation of the call graph based on domain-specific knowledge (such as the bytecode redirection, and windowing system methods emptying) is sound if and only if all cases are envisioned. Given the complexity and scale of a framework such as Android, this completeness is hard to prove.
For Spark-based analysis, the analysis is sound if and only if the object and static fields are correctly initialized.
Hence the analysis may be sound for some entry-points and unsound for others. 
For a framework such as Android, there is no oracle for formally answering those questions.
However, for those entry points when the CHA-based results and the Spark-based results are identical it is a strong piece of evidence of soundness.
For the rest, comparison with documentation or runtime data is required.
Finally, our results hold as far as there is no serious bug in the implementation of any part of the static analyses (e.g., entry point initialization and bytecode redirection), as well as in the glue and measurement code we wrote.
 
\subsection{The Impact of Service Identity Inversion}
\label{sec:serviceIdentityInversion}
\makeatletter{}A legitimate question to ask is whether or not service identity inversion has an impact on the resulting permission set.
To answer that, we ran \sparkIntelligent{} with and without activating service identity inversion.
Within the set of entry points which did not time out, two have a bigger permission set when service identity inversion is turned off.
For instance, method \method{<android.net.ConnectivityManager boolean requestRouteToHost(int,int)>} has one more permission \perm{CONNECTIVITY\_INTERNAL} when service inversion is disabled. 
This permission is not required for the entry point according to the official documentation\footnote{\url{http://developer.android.com/reference/android/net/ConnectivityManager.html}} which validates the usefulness of the service identity inversion building block.
Service inversion may only impact a few entry points but not taking it into account leads to wrong permission sets. 
\subsection{Limitations}
\makeatletter{}
\subsubsection{Native Code}
The Android framework is a real-world large-scale framework, featuring heterogeneous layers written in different languages. 
For Android 2.2 most Android permissions (126/134) are checked in the Android Java framework only. 
Our approach is complete for these 126  permissions, but incomplete for  the eight permissions checked in native C/C++ code. 
These eight permissions are: \perm{BLUETOOTH\_ADMIN}, \perm{BLUETOOTH}, \perm{INTERNET}, \perm{CAMERA}, \perm{READ\_LOGS},  \perm{WRITE\_EXTERNAL\_STORAGE}, \perm{ACCESS\_CACHE\_FILESYSTEM} and \perm{DIAGNOSTIC}.
\subsubsection{Reflection in the Framework}
 
If the framework uses reflection, then the call graph construction is incomplete by construction.
Fortunately, the Android framework uses reflection in only 7 classes. We manually analyzed their source code.
Five of those classes are debugging classes. The \class{View} class uses reflection for handling animations. 
Finally, the \class{VCardComposer} uses reflection in a branch that is only executed for testing purpose. 
In all cases, the code is not related to system resources hence no permission checks are done at all. 
This does not impact the static analysis of the Android framework.
\subsubsection{Dynamic Class Loading}
The Java language has the possibility to load classes dynamically. 
Static analysis cannot deal with this since the loaded classes are only known at runtime. 
We found that eight classes of the Android system are using
the \method{loadClass} method. After manual check, six of them are system management classes and are either not linked to permission checks (ex: instrumenting an application)
or have to be accessed through a service.
Two are related to the \package{webkit} package. 
They are used in the \class{LoadFile} and \class{PluginManager} classes. 
In both cases, permissions are checked \emph{before} loading classes, and not inside the loaded classes. 
Thus, there is no missed permission enforcement point either.
\subsubsection{Spark}
 
Our model of the Android framework focuses on services and missed the initialization of other Android components (e.g., content providers).
In other words, Spark is sound with regards with our model of Android components.

\section{Computing Permission Gaps}\label{sec:permission-gap}
  \makeatletter{}
We now have static analyses to compute the mapping between Android API methods and their required permissions. 
This section first presents a method to efficiently compute the required permission set and the corresponding permission gap (permissions declared but not used), if any.
Then we present the results of an empirical study that show the existence of permission gaps in the wild.
\subsection{A Calculus for Permission Analysis}
\label{subsec:def}
This section describes the permission gap inference as a calculus on top of a boolean matrix algebra.
Permission inference is at heart a reachability analysis (does the application reach a permission check?), the goal of this calculus is to "factorize" the static analysis, so as to be much more efficient.
Let $app$ be an application. The \emph{access vector} for $app$ is a boolean vector $AV_{app}$ representing the entry points of the framework under study.  Thus, the length of vector $AV$ is the number of entry points of  framework $\mathcal{F}$. An element of the vector is set to \emph{true} if the corresponding entry point is called by the application. Otherwise it is set to \emph{false}.
Let us consider a framework with four entry points ($e_1$, $e_2$, $e_3$, $e_4$), and an application $foo$ that reached $e_1$, $e_2$ and $e_3$ but not $e_4$. $AV_{app}$ reads:
$$ AV_{foo} = \left(1, 1, 1, 0\right) $$ 
We define the \emph{permission access matrix} $M$ as a boolean matrix which represents the relation between entry points of the framework and permissions. 
The rows represent entry points of the framework and the columns represent permissions. 
A cell $M_{i,j}$ is set to \emph{true} if the corresponding entry point (at row $i$)  accesses a resource protected by the permission represented by column $j$. Otherwise it is set to \emph{false}.
For a framework with four entry points ($e_1$, $e_2$, $e_3$ and $e_4$) 
and three permissions ($p_1$, $p_2$ and $p_3$), the permission access matrix reads:
$$ M = 
\bordermatrix{
  & p_1	& p_2 & p_3  \cr
e_1 & 1 & 0 & 0  \cr
e_2 & 1 & 0 & 0  \cr
e_3 & 0 & 0 & 0  \cr
e_4 & 0 & 1 & 0  \cr
}
$$
\ldots meaning that $e_1$ and $e_2$ require permission $p_1$, $e_3$ requires no permission and  $e_4$ requires permission $p_2$.
Let $app$ and $\mathcal{F}$ be an application and a framework respectively.  
The inferred permissions vector, $IP_{app}$, is a boolean vector representing the set of inferred permissions for application $app$. By using the boolean operators AND and OR instead of arithmetic multiplication and addition in the matrix calculus, we have:
$$IP_{app} = AV_{app} \times M$$
A cell $IP_{app}(k)$ equals to \emph{true} means that the permission at index $k$ is required by $app$.   Using $AV_{app}$ and $M$ from the  previous examples, the inferred permissions vector for $app$ is:
\begin{eqnarray*}
IP_{app} & = &  \left( \begin{array}{ccccc} 1 & 1 & 1 &  0 \end{array}\right) \cdot
\left( \begin{array}{ccc}
1 & 0 & 0 \\
1 & 0 & 0 \\
0 & 0 & 0 \\
0 & 1 & 0 \end{array} \right) \\
IP_{app} & = & \left( \begin{array}{ccc} 1 & 0 & 0  \end{array}\right)
\end{eqnarray*}
\ldots meaning that the application should declare and only declare permissions $p_1$.
\subsection{Extraction of $M$ and $AV$}
The permission access matrix $M$  is based on a static analysis of framework $\mathcal{F}$. 
As shown in Section \ref{sec:application-on-android}, we first compute a call graph for every entry point of the framework and then detect whether or not permission checks are present in the call graph. 
A permission enforcement point (PEP) is a vertex of a  call graph whose signature corresponds to a system  method that checks permission(s). Each PEP is associated with a list of required permissions $perms_{PEP}$.
Matrix $M$ is constructed as follows:
it is a matrix of size  (|entry points| $\times$ |high level permissions|); all elements of $M$ are initialized to false; for each $e_i$ that reaches one or more PEP,  and for each permission $j$ in $perms_{PEP}$, $M(i,j) = true$. 
In other terms, \emph{M is a condensed version of the reachability information that is latent in call graphs.}
Let us take the example of Figure \ref{fig:generic-application-framework} in Section \ref{sec:manifest}.
It shows a framework with four entry points $(e_1,e_2,e_3, e_4)$, and three permissions $(p_1,p_2,p_3)$.
For every of those entry points a call graph is constructed. 
Three of those call graphs have a PEP node: $e_1$ and $e_2$ have PEP $ck_1$ which checks permission $p_1$ and $e_4$ has PEP $ck_2$ which checks permission $p_2$.
On the figure a dashed arrow connects each PEP to the permission(s) it checks.
The framework matrix is then matrix $M$ presented above (see Section \ref{subsec:def}).
Extracting $AV$ simply means listing the list of entry points of a framework $\mathcal{F}$ called by an application $app$. The application example in Figure \ref{fig:generic-application-framework} uses a single entry point, and $ AV_{ex} = \left(1, 1, 1, 0\right) $.
\subsection{Computing the Permission Gap }
The permission gap is the difference between the permissions extracted from $IP_{app}$ and the declared permissions \pdeclared.
In Figure \ref{fig:generic-application-framework}, using matrix $M_{ex}$ and vector $AV_{ex}$ of the example framework and application, we obtain a list of inferred permissions only containing $p_1$.
If the application declares $p_1$ and $p_2$, the permission gap is $\{p_2\}$.
\makeatletter{}
We ran our tool on two datasets of Android applications. The first comes from an alternative Android Market\footnote{\url{http://www.freewarelovers.com/android}} and contains 1329 android applications.
For the second one, we consider the top 50 downloaded applications of all 34 top-level categories of the Official Android Market, as well as the top 500 of all applications and the top 500 of new applications (on February, 23$^{rd}$ 2012).
After removal of duplicates (the applications appearing in several rankings), the second dataset contains 2057 applications.
 
\textbf{Alternative Android Market:} 
We discard 587 applications that use reflection and/or class loading. Of the 742 remaining applications, 94 are declaring one or more permissions which they do not use.
Consequently, \emph{we identify a permission gap for 94 Android applications}.
We define the ``area of the attack surface'' with respect to permission gaps, as the number of unnecessary permission.
In all, among applications suffering from a permission gap, 
76.6\% have an attack surface of 1 permission,
19.2\% have an attack surface of 2 permissions, 2,1\% of 3 permissions and also 2,1\% of 4 permissions.
\textbf{Official Android Market:} 
We discard 1378 applications that use reflection and/or class loading. On the 679 remaining applications, 124 are declaring one or
more permissions which they do not use.
In all, among applications suffering from a permission gap, 
64.5\% have an attack surface of 1 permission,
23.4\% have an attack surface of 2 permissions, 12.1\% of 3 or more permissions.
To sum up, those results show that permission gaps exists, and that our approach allows developers to fix the declared permission list in order to reduce the attack surface of permission-based software.

\section{Related Work}\label{sec:related-work}
	\makeatletter{}
We have presented an approach to reduce the attack surface of permission-based software.
The concept of ``attack surface'' was introduced by Manadhata and colleagues \cite{Manadhata2011}, it describes all manners \emph{in which an adversary can enter the system and potentially cause damage}. This paper describes a method to identify the attack surface of Android applications, which is an important research challenge given the sheer popularity of the Android platform.
In the context of Android, reducing the attack surface is adhering to the principle of least privileges introduced by Saltzer \cite{Saltzer1975}.
\subsection{On the Java Permission Model}
While the Android permission model is different from the one implemented in Java, the following pieces of research present related and relevant points to put our contribution in perspective.
Koved and al. described a new static analysis \cite{Koved:2002:ARA} to generate a permission list for a Java2 program (in the Java permission model).  
Geay et al. \cite{conf/icse/GeayPTRD09} presented an improved methodology.
We also use static analysis but in the context of Android which differs from a Java environment especially with respect to the binder mechanism linking Android API to services. As shown in our evaluation, the binder prevents off-the-shelf Java static analysis tools to resolve remote call to a service.\newline
Pistoia et al. \cite{conf/ecoop/PistoiaFKS05} presented a static analysis to identify portions of the
code which should be made privileged. 
This issue does not arise in the Android framework since code is not privileged per se, the access control is instead done at entry points. This means that the Android framework designers must be careful of creating unique entry points protected by permission enforcement points, but does not impact our static analysis.\newline
Centonze et al. \cite{CeNaFiPi2006} analyzed role-based access control (RBAC) mechanisms using static analysis. 
When a protected operation manipulates data, this data should not be directly or indirectly accessible by a path not defined in the policy.  If not, the operation is said to be \emph{location-inconsistent}. 
The tool they developed can check  whether or not an RBAC policy for JavaEE programs is location consistent or present some flaws. The Android system defines permissions which
protect operation which in turn manipulate protected data. Our goal consists of computing permission gaps which may reveal a violation of the principle of least privilege. Whether Android protected operations are location consistent is out of scope of this paper.\newline
Also related to role-based access control, Pistoia et al. \cite{conf/icse/PistoiaFFY07} formally model RBAC and statically check the consistency of a JavaEE based RBAC system. We check that permission lists of Android applications respect the principle of least privilege. 
The concepts are the same (Android permissions could be approximated to roles, and we check which roles are needed at every point of the Android framework) but the target systems are not. 
 Interestingly, we use a similar approach for solving the Binder problem as they do for solving the remote method invocation problem: instead of statically analyzing the Binder/RMI code which would not resolve the method, a mapping is computed from the call to a remote method to the remote method itself. A major difference though is that in the case of Android system services and context must be initialized beforehand to simulate a correct system state.
\subsection{On the Android Permission Model}
The Android security model has been described as much in the gray literature \cite{Enck2009,Shabtai2009} as in the official documentation \cite{android}.
Different kinds of issues have been studied such as social engineering attacks \cite{Hoffman2011}, collusion attacks \cite{Marforio2011}, privacy leaks \cite{Gibler2011} and privilege escalation attacks \cite{Felt2011b,davi2011privilege}.
In contrast, this paper does not describe a particular weakness but rather a software engineering approach to reduce potential vulnerabilities.\newline
However, we are not describing a new security model for Android as done by  \cite{Nauman2010,Ongtang2011,quire2011,Conti2011,Bugiel2011}. For instance, Quire \cite{quire2011} maintains at runtime the call chain and data provenance of requests to prevent certain kinds of attacks.
In this paper, we do not modify the existing Android security model and we devise an approach to mitigate its intrinsic problems.\newline
Also, different authors empirically explored the usage of the Android model.
For instance, Barrera et al. \cite{DBLP:conf/ccs/BarreraKOS10} presented an empirical study on how permissions are used. In particular, they used visualizing techniques such as self-organizing maps to identify patterns of permissions depending on the application domain, and patterns of permission grouping. Other empirical studies include Felt's one \cite{Felt2011} on the effectiveness of the permission model, and Roesner's one \cite{Roesner2011} on how users react to permission-based systems.
While our paper also contains an empirical part, it is also operational because we devise an operational software engineering approach to tame permission-based security models in general and Android's one in particular.\newline
Enck et al \cite{Enck2009a} presented an approach to detect dangerous permissions and malicious permission groups. They devised a language to express rules which are expressed by security experts. Rules that do not hold at installation time indicate a potential security problem hence a high attack surface. Our goal is different: we don't aim at identifying risks identified from experts, but to identify the gap between the application's permission specification and the actual usage of platform resources and services. Contrary to \cite{Enck2009a}, our approach is fully automated and does not involve an expert in the process.\newline
PScout \cite{au2012pscout} is a static analysis designed concurrently with our work.
It also uses Soot but only relies on CHA and does not use Spark.
Our works compares and validates part of their results in Section \ref{sec:comparison-pscout}.
Finally, Felt et al. \cite{Felt2011a} concurrently worked on the same topic as this paper. They published a very first version of the map between developer's resources (e.g., API calls) and permissions. 
Interestingly, we took two completely different approaches to identify the map: while they use testing, we use static analysis. As a result, our work validates most of their results although we found several discrepancies that we discussed in details in Section \ref{sec:comparison-felt}. But the key difference is that our approach is fully automated while theirs requires manually providing testing ``seeds'' (such as input values). However, in the presence of reflection, their approach works better if the tests are appropriate. Hence, we consider that both approaches are complementary, both at the conceptual level for permission-based architectures, and concretely for reverse-engineering and documenting Android permissions.
Mustafa et al. \cite{MustafaTechReport2012} worked on the analysis of system services. Their approach is to extract a sub call graph using a context-sensitive backward slicing method starting from permission check methods. Their analysis is more precise since they capture conditions under which permissions are checked. However, they only consider independent system services and do not handle RPC. We, on the other hand, start the analysis from the Android API entry points and handle services RPC links.
   
\section{Conclusion}\label{sec:conclusion}
	\makeatletter{}
In this paper, we have empirically demonstrated that off-the-shelf static analysis can not address the extraction of permissions in Android.
At least three static analysis components must be put together in order to use Class Hierarchy Analysis (CHA) and field-sensitive static analysis (Spark) for analyzing Android’s permissions.
Those are (1) a string analysis, (2) service identity inversion and (3) entry point and service initialization for Spark.
 
We have compared our work with PScout \cite{au2012pscout} and Felt \cite{Felt2011a}.
We show that our approach confirms results from concurrent work and that static analysis is complementary to dynamic analysis.
Moreover, we have presented a generic approach to reduce the attack surface
of permission-based software\footnote{details at \paperurl{}} .
We have extensively discussed the problematic consequences of having more
permissions than necessary and showed that the problem can be mitigated using
static analysis.
The approach has been fully implemented for Android, a permission-based platform for mobile devices. 
For end-user applications, our evaluation revealed that 94/742 and 35/679 applications crawled from Android application stores indeed suffer from permission gaps. 
The security architecture of permission based software in general and Android in particular is complex. In this paper, we abstracted over several characteristics of the platform such as low-level permissions.
We are now working on a modular approach that would be able to analyze native code and bytecode in concert and to combine the permission information from both.
Furthermore, we are exploring how to express permission enforcement as a cross cutting concern, in order to automatically add or remove permission enforcement points at the level of application or the framework, according to a security specification.
 
\bibliographystyle{abbrv}
\bibliography{bib/bib}  
\section{Acknowledgments}
This research is supported by the National Research Fund, Luxembourg (AFR grant 1081630). We also would like to thank Eric Bodden for his help in using the Soot analysis toolkit.
\makeatletter{}\begin{biography}[{\includegraphics[width=1in, height=1.25in, clip, keepaspectratio]{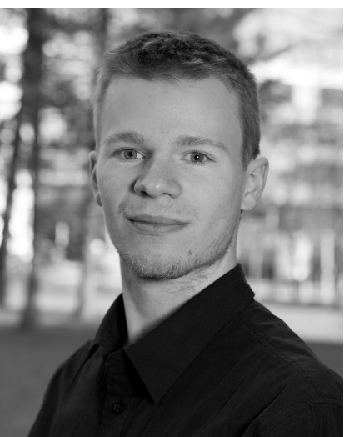}}]{Alexandre Bartel}
Alexandre Bartel is a PhD candidate in Software Engineering at the University of Luxembourg at the Interdisciplinary Center for Security Reliability and Trust (SnT) / Serval Team. 
He received a M.S. degree in Computer Engineering from INPG-Esisar, France in 2010 and a M.S. from KTH, Sweden in 2010.
His current research focuses on analyzing permission-based software and bytecode monitoring for in-vivo smartphone security. 
\end{biography}
\begin{biography}[{\includegraphics[width=1in, height=1.25in, clip, keepaspectratio]{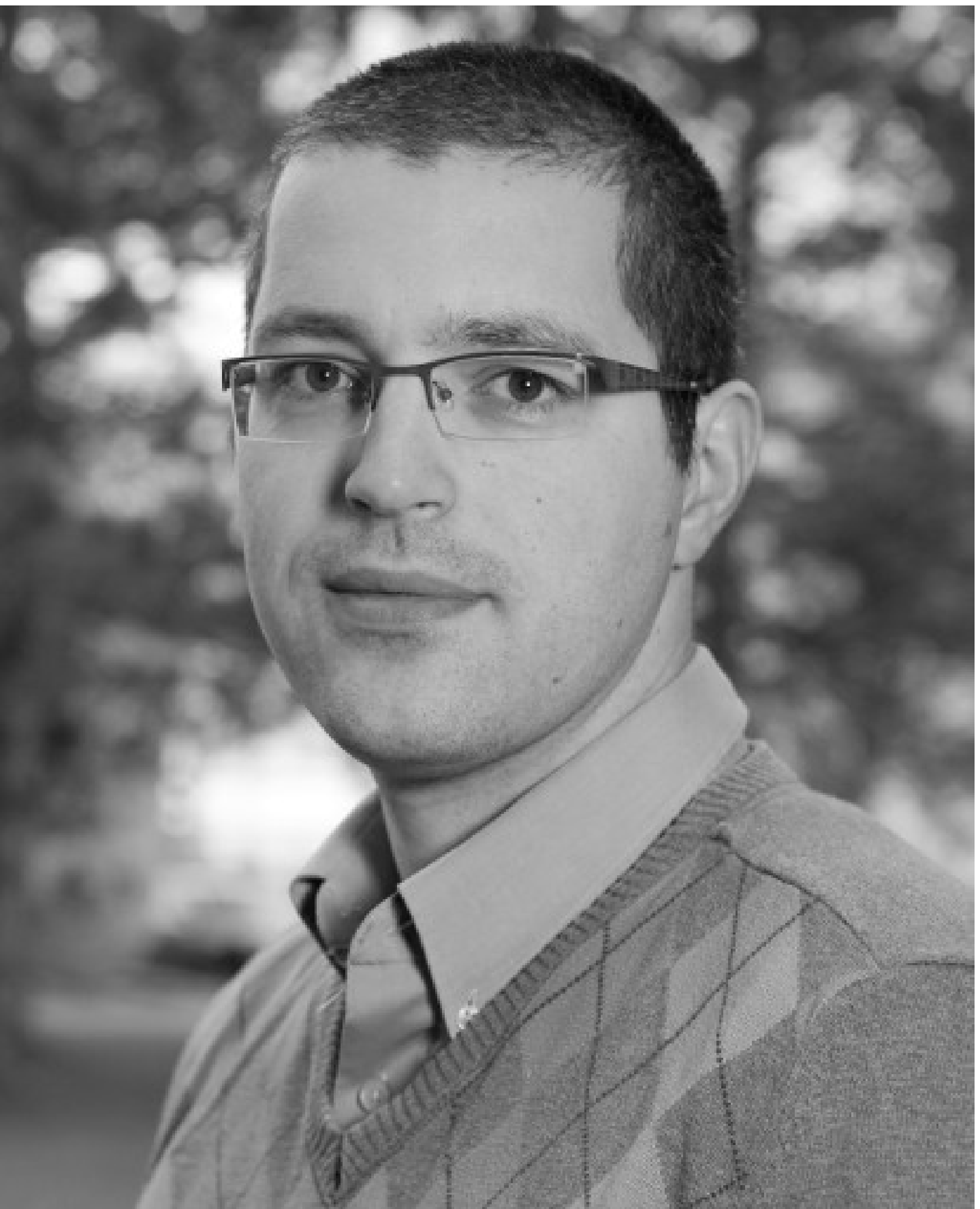}}]{Jacques Klein}
Jacques Klein is research scientist at the University of Luxembourg and at the Interdisciplinary Centre for Security, Reliability and Trust (SnT). 
He received a Ph.D. degree in Computer Science from the University of Rennes, France in 2006. 
His main areas of expertise are: Software Security by applying software engineering to security; Model-Driven Engineering, with a focus on model composition and model@runtime; Software Testing, and Software Product Lines. 
\end{biography}
\begin{biography}[{\includegraphics[width=1in, height=1.25in, clip, keepaspectratio]{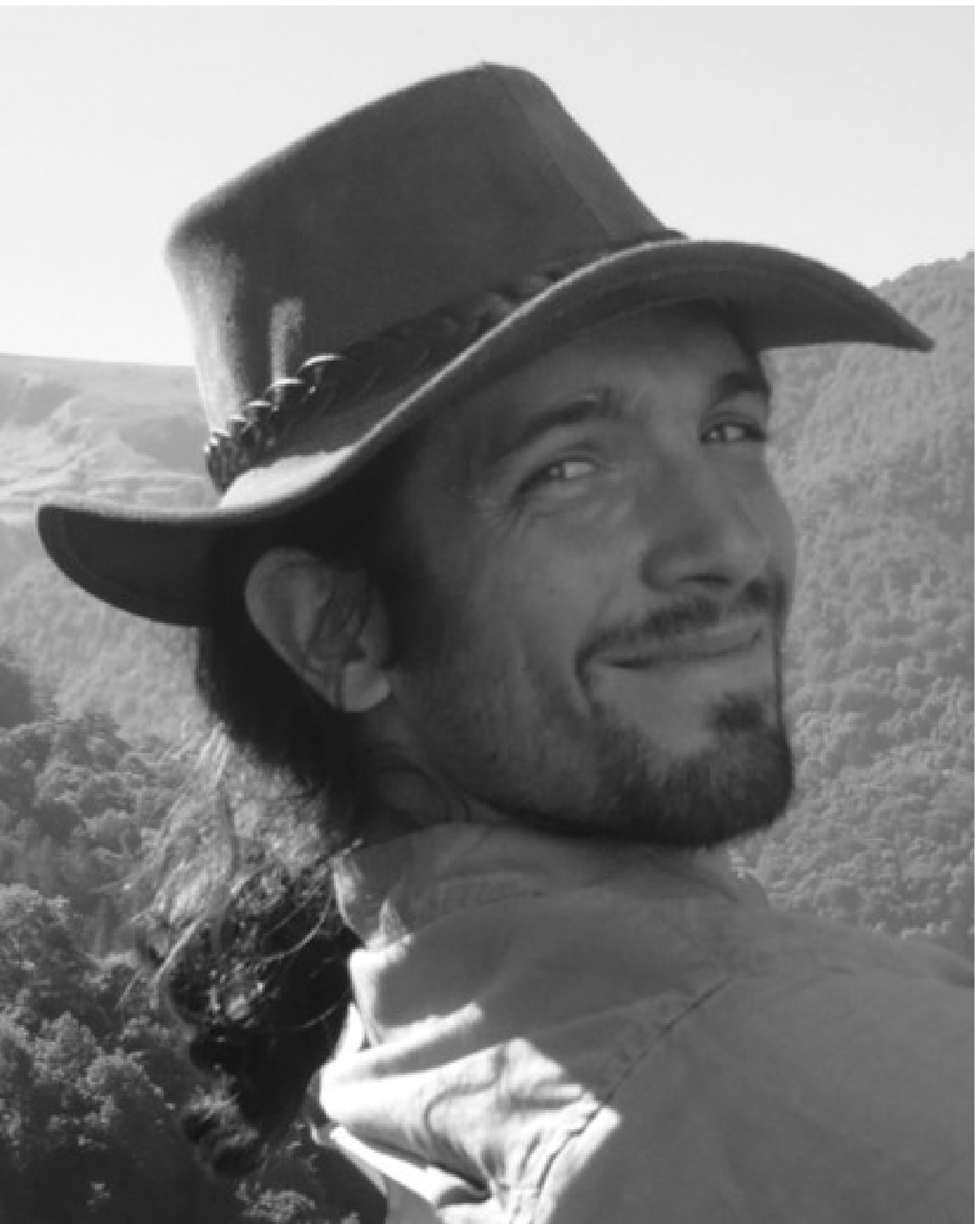}}]{Martin Monperrus}
Martin Monperrus has been an associate professor at the University of 
Lille since 2011. He was previously with the Darmstadt University of 
Technology as a research associate. He received a Ph.D. from the 
University of Rennes in 2008 and a Master's degree from the Compiègne 
University of Technology in 2004.
\end{biography}
\begin{biography}[{\includegraphics[width=1in, height=1.25in, clip, keepaspectratio]{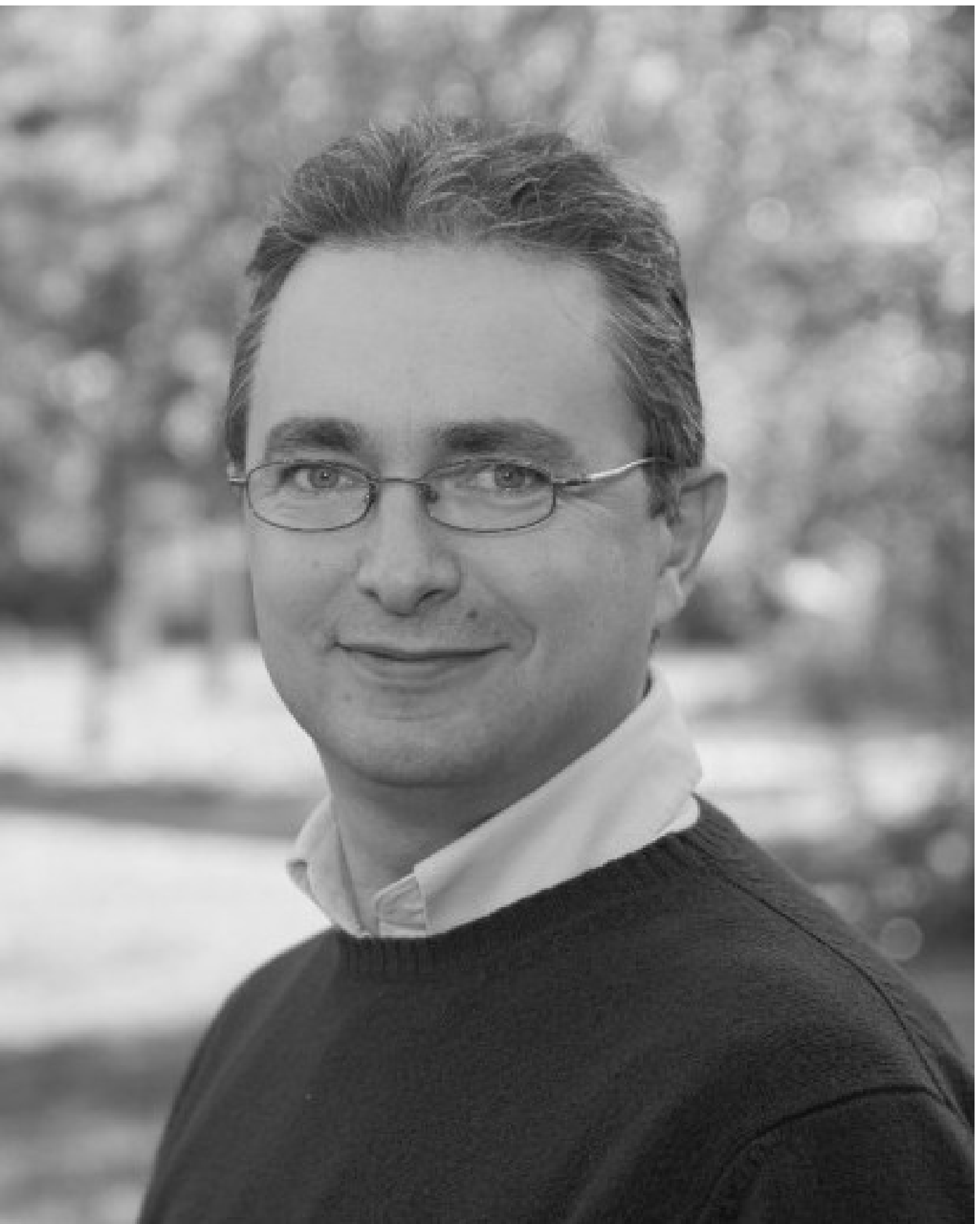}}]{Yves Le Traon}
Yves Le Traon is professor at the University of Luxembourg, in the domain of software engineering, reliability, testing and security. 
His current research interests include and combine Software Product Line re-engineering and testing, Android security, mutation testing, model-driven security, and SBSE. 
He received a Ph.D. degree in Computer Science at the “Institut National Polytechnique” in Grenoble, France, in 1997. 
He was associate professor in France at the University of Rennes, and then full professor at Telecom Bretagne until he reaches Luxembourg in 2009. 
He is currently the head of the CSC Research Unit (Department of Computer Science) and an active member of the Interdisciplinary Centre for Security, Reliability and Trust (SnT), where he leads the SERVAL group (SEcuRity design and VALidation). 
He has been on the program, steering, or organization committees of many international IEEE software engineering conferences. 
He also belongs to the steering committee of IEEE ICST. He is a member of the IEEE Computer Society.
\end{biography}
\vfill 
\end{document}